\documentstyle[amsfonts,eqsecnum,multicol,fancyheadings,prd,aps]{revtex}

\begin{document}
\draft

\title{Dimensional reduction and gauge group reduction in\\ 
Bianchi-Type cosmology}

\author{J.\ M.\ Pons
\footnote[1]{Electronic address: pons@ecm.ub.es}}
\address{Departament d'Estructura i Constituents de la Mat\`eria, 
Universitat de Barcelona, \\
and Institut de F\'\i sica d'Altes Energies,\\
Diagonal 647, E-08028 Barcelona, Catalonia, Spain}

\author{L. C. Shepley
\footnote[2]{Electronic address: larry@helmholtz.ph.utexas.edu}}
\address{Center for Relativity, Physics Department, \\
The University of Texas, Austin, Texas 78712-1081, USA\\~}

\date{Submitted 5 February 1998}
\maketitle
\begin{abstract}

In this paper we examine in detail the implementation, with its 
associated difficulties, of the Killing conditions and gauge fixing 
into the variational principle formulation of Bianchi-Type 
cosmologies.  We address problems raised in the literature concerning 
the Lagrangian and the Hamiltonian formulations: We prove their 
equivalence, make clear the role of the Homogeneity Preserving 
Diffeomorphisms in the phase space approach, and show that the number 
of physical degrees of freedom is the same in the Hamiltonian and 
Lagrangian formulations.  Residual gauge transformations play an 
important role in our approach, and we suggest that Poincar\'e 
transformations for special relativistic systems can be understood as 
residual gauge transformations.  In Appendices, we give the general 
computation of the equations of motion and the Lagrangian for any 
Bianchi-Type vacuum metric and for spatially homogeneous Maxwell 
fields in a nondynamical background (with zero currents).  We also 
illustrate our counting of degrees of freedom in an Appendix.

\end{abstract}

\pacs{04.20.Fy, 11.10.Ef, 98.80.Hw \hfill gr-qc/9805030}

 \setlength{\columnseprule}{0pt}\begin{multicols}{2}


\section{Introduction}
\label{sec:intro}

A spatially homogeneous cosmological model is a manifold $\cal M$ with 
a Lorentzian metric tensor {\bf g} invariant under a group of 
isometries whose three-dimensional, spacelike, invariant hypersurfaces 
foliate $\cal M$.  In the models we will treat, this group is 
generated by three spacelike vector fields ${\bf K}_{a}$, which span a 
Lie algebra defined by their commutation relations:
\begin{equation}
	[{\bf K}_{a},{\bf K}_{b}] = C^{c}_{ab}{\bf K}_{c} \ .
	\label{1.com}
\end{equation}
The invariance of the metric is expressed by the vanishing of its Lie 
derivatives with respect to these vectors: 
${\cal L}_{\bf K_{a}}{\bf g}=0$.  
The structure constants are antisymmetric in their lower indices and 
obey the Jacobi relation:
\begin{equation}
	C^{c}_{ab} = - C^{c}_{ba}\ ;\
	C^{a}_{be}C^{e}_{cd} + C^{a}_{ce}C^{e}_{db}
	    + C^{a}_{de}C^{e}_{bc}
	    = 0 \ .
	\label{1.anti}
\end{equation}
The Bianchi classification of these algebras into nine Types  
(see \cite{taub51}) is the source of the term Bianchi-Type cosmology.  
In three dimensions the Jacobi relation is equivalent to
\begin{equation}
	C^{a}_{ea}C^{e}_{bc} = 0\ .
	\label{1.jacobi}
\end{equation}
The algebras with vanishing $C^{a}_{ea}$ are called Class A and those 
with nonvanishing $C^{a}_{ea}$ are called Class B 
\cite{ellis/maccalum69}.

In a suitable basis, {\bf g} can be represented by components 
$g_{\mu\nu}$ which depend only on a single variable, cosmic time $t$.  
The Einstein field equations are (coupled, non-linear) ordinary 
equations.  (The choice of this basis is part of this paper.)  The 
Einstein-Hilbert Lagrangian density for the metric may be computed in 
terms of these components and their $t$-derivatives.  It is well known 
that in Class A models this reduced Lagrangian does correctly 
reproduce the field equations, but it does not necessarily do so in 
Class B models 
\cite{hawking69,maccallum71,maccallum/taub72,ryan74,sneddon76} because 
of a spatial divergence that is automatically zero only in Class A 
\cite{maccallum79}.

Bianchi-Type cosmology is still a subject of debate and --- we 
think --- some misunderstanding.  It has been claimed, surprisingly, 
that the number of degrees of freedom in the Lagrangian and 
Hamiltonian formulations of these models do not agree 
\cite{ashtekar/samuel91}.  This result would mean that the two 
formulations are not physically equivalent.  Another issue, related to 
this one, poses the question as to what is to be considered a gauge 
transformation, and therefore a redundancy in the physical description 
in these models.

In this paper, we will address three subjects: the Lagrangian and 
Hamiltonian formulations of Bianchi-Type models (dimensional 
reduction), the concept of gauge freedom for these models, and the 
fact that the number of degrees of freedom in the Lagrangian and 
Hamiltonian formulations do agree.  We do not solve the first subject, 
which remains a problem for Class B models, although we clarify some 
points that are also relevant for Class A models.  From our analysis 
we give an answer to the second and third problems.  We show that the 
gauge freedom is dictated by the diffeomorphism invariance of the 
original theory and leads to the physical equivalence of the 
Lagrangian and Hamiltonian formulations.

We treat homogeneous models in general relativity as constrained 
dynamical systems.  In particular we clarify the two aspects that 
merge when we move from the superspace of all metrics to the 
minisuperspace of homogeneous metrics.  On the one hand there is the 
dimensional reduction from $3+1$ spacetime dimensions to one time 
dimension.  This dimensional reduction comes from the symmetry 
associated with the Lie algebra of the Killing vectors.  On the other 
hand there is the elimination of gauge degrees of freedom.  These two 
types of reduction present different types of problems, which we 
analyze, but both are necessary in order to reduce the original gauge 
group of four-diffeomorphism invariance to the gauge group of 
time reparameterization invariance.

This reduction procedure will be undertaken in four steps.  As an 
outcome of our analysis, a) we will clarify some points concerning the 
Lagrangian and Hamiltonian reduced formulations; b) we will be able to 
point out clearly what is the gauge group for these models; and c) we 
will show that the correct reduction procedures in both the Lagrangian 
and Hamiltonian formulations lead to a number of degrees of freedom 
which is always the same in either formulation.  The first step, in 
Section \ref{sec:1step}, consists of adapting time coordinates to the 
symmetry group.  The second step, in Section \ref{sec:2step}, is to 
adapt spatial coordinates to the group by adopting a gauge in which 
the shift vector depends only on time.  The third step, in Section 
\ref{sec:3step}, is to adopt a gauge in which the shift vector 
vanishes.  Finally, the fourth step, in Section \ref{sec:4step}, is to 
eliminate residual three-diffeomorphisms.  After each step we examine 
the status of what remains of the gauge group.  In Section 
\ref{sec:zeroshift} we examine the loss of constraints caused by the 
gauge fixing.  We perform, in Section \ref{sec:hamiltonian}, the 
Hamiltonian analysis, which is presented in a summary form because it 
is parallel to the Lagrangian analysis.  Section \ref{sec:conclusion} 
is devoted to conclusions.

In Appendix \ref{app:homog} we present explicit equations for 
spatially homogeneous metrics in full detail, including general lapse 
function and shift vector, for any Bianchi Type, where coordinates 
have been chosen so that all variables depend only on time.  The 
purpose is to provide concrete examples for our formalism.  A simpler 
example, which also can serve to illustrate our methods, is provided 
in Appendix \ref{app:maxwell}: It is the case of a spatially 
homogeneous electromagnetic field in a spatially homogeneous 
background metric.  Appendix \ref{app:varprin} reviews the general 
literature on the relationship between setting gauge conditions and 
variational principles.  It gives justification to some results used 
in Section \ref{sec:3step} and Section \ref{sec:hamiltonian}.  
Appendix \ref{app:freedom} illustrates our counting of degrees of 
freedom in the Bianchi Type I case and shows how spacetime rigid 
symmetries may be considered as residual gauge transformations.


\section{Time}
\label{sec:1step}

We start with the general setting for Bianchi models: Call $\Bbb T$ 
the set, each element of which is a metric tensor {\bf g} and three 
vector fields ${\bf K}_{a}$ whose orbits generate three-dimensional 
hypersurfaces which foliate the four-dimensional manifold $\cal M$; 
the Lie derivatives of {\bf g} with respect to the ${\bf K}_{a}$ 
vanish.  The invariance of {\bf g} and even the definition of the 
${\bf K}_{a}$ need only be locally defined for much of what we do; if 
the vector fields are globally defined, we speak of global homogeneous 
cosmologies \cite{ashtekar/samuel91}.

The diffeomorphisms on $\cal M$ can be viewed in an active sense, 
mapping points onto other points.  In that case, an element (metric 
plus three vector fields) of $\Bbb T$ will in general be mapped to 
another element in $\Bbb T$.  If there is a covering of $\cal M$ by 
coordinate patches, each diffeomorphism may also be viewed in a 
passive sense as a collection of coordinate transformations.  It is 
this passive sense which is closer to the physical principle which is 
a motivation for General Relativity, that physics should not depend on 
coordinates.  It is, in fact, somewhat easier to adopt the language of 
a metric being determined by its components in a coordinate patch; a 
diffeomorphism is then a transformation which preserves the metric 
{\bf g} and the vectors ${\bf K}_{a}$ but in general changes their 
components $g_{\mu\nu},K_{a}\!^{\mu}$.  Nevertheless, we shall adopt 
the active view when convenient.  (Greek indices range over 
0,1,2,3, with the coordinate $x^{0}$ being the time $t$.  Latin 
indices will range over 1,2,3.)

Given a coordinate neighborhood $\cal N$, we consider all metric 
tensors, defined by their components $g_{\mu\nu}$ in $\cal N$, which 
are invariant under isometries defined by the ${\bf K}_{a}$ and in 
which the ${\bf K}_{a}$ are spacelike.  The metric components satisfy 
the Killing equation:
\begin{equation}
    0 = \left({\cal L}_{\bf K_{a}}{\bf g}\right)_{\mu\nu}
        = K_{a}\!^{\sigma}g_{\mu\nu,\sigma}
        + g_{\sigma\nu}K_{a}\!^{\sigma}\!_{,\mu}
        + g_{\mu\sigma}K_{a}\!^{\sigma}\!_{,\nu}\ ,
            \label{killeq}
\end{equation}
where comma denotes partial differentiation.  We can consider $\Bbb T$ 
as being the collection of all such tensor components, each element of 
$\Bbb T$ being the collection of metric and Killing vector components, 
the coordinate system being understood.  Any of the elements of 
$\Bbb T$ foliates $\cal M$ by three-dimensional space-like homogeneous 
hypersurfaces, namely the integral surfaces of the Killing vector 
fields.

A change of coordinates --- a diffeomorphism --- will in general 
change the form of the metric components.  In that sense, the 
diffeomorphism group ${\bf Diff}({\cal M})$ on $\cal M$ is realized as 
a group {\bf D} acting on $\Bbb T$.  The physics doesn't change, of 
course, under a coordinate transformation, and symbolically we can 
write
\[ 
    Physics\ (of\ a\ Bianchi\ Type)\ 
    = {\Bbb T} / {\bf D} \ .
\]

These conditions need only hold locally for much of what we will be 
doing.  In that sense, we really are working with a Lie algebra of 
infinitesimal isometries rather than a Lie group, though we will 
continue to speak of the isometries as a symmetry group.  When we need 
to require global homogeneity, we will clearly specify so.

The existence of these isometries of the metric can be used to 
simplify greatly the equations of motion, which we take to be the 
vacuum Einstein equations.  At this point, the full gauge group 
{\bf D} of General Relativity, generated by ${\bf Diff}{\cal M}$, is 
still operating in ${\Bbb T}$.  We will be looking for a reduced 
Lagrangian, with the Killing conditions built in, capable of 
describing these Bianchi models.  We do so in four steps: adapting 
time, adapting space, fixing the gauge by requiring the shift vector 
(to be defined later) to vanish, and fixing the residual gauge 
freedom.

We now proceed to the first step, which will partially fix the gauge 
(that is fixing coordinates) by concentrating on the three-dimensional 
homogeneous hypersurfaces.  Suppose that the homogeneous hypersurfaces 
happen to coincide with the hypersurfaces $t=\rm const$ in the 
coordinates defining an element of $\Bbb T$, 
$(g_{\mu\nu},K_{a}\!^{\sigma})$; then we introduce an equivalence 
relation by saying that another element of $\Bbb T$ is equivalent to 
this one if it is produced by an element of {\bf D}, that is by a 
diffeomorphism.  The set of these equivalence classes is 
${\Bbb T}_{t}$, and the full group {\bf D} has been reduced to a 
smaller group ${\bf D}_{t}$, namely those diffeomorphisms which 
preserve the condition that the homogeneous hypersurfaces are defined 
by $t=\rm const$.

In this first step, we have made a choice of spacetime local 
coordinates such that the surfaces $\Sigma_t$ of constant time 
coincide with the foliation defined by the Killing vectors.  The time 
coordinate is a function whose curl vanishes under Lie differentiation 
by any of the Killing vectors:
\begin{equation} 
    {\cal L}_{{\bf K}_a} {\bf d}t = 0\ .	
        \label{adapt}
\end{equation}
Our Killing vectors now take the general form 
(where $\bbox{\partial}_{i}\equiv\partial/\partial x^{i}$)
\begin{equation} 
    {\bf K}_a = K_a\!^i(t,x) \bbox{\partial}_{i}\ . 
\end{equation}
Notice the possible time dependence of ${\bf K}_a$.  (We 
denote the four spacetime coordinates $t,x^{i}$ by $t,x$ when 
they are used as the arguments of a function.)
 

\subsection{Gauge group after the first step}
\label{subsec:1step.gauge}

In these spacetime coordinates $\{t,x^{i}\}$, a general infinitesimal 
reparameterization (coordinate transformation) is generated by a 
vector field $\epsilon^\mu(x)\bbox{\partial}_\mu$:
\begin{equation}
    \pmatrix{t \cr x^{i}} \longrightarrow
    \pmatrix{t+\delta t\cr x^{i}+\delta x^{i}}
    = \pmatrix{t +\epsilon^{0} \cr x^{i} +\epsilon^{i}}\ .
\end{equation}
To preserve equation (\ref{adapt}), that is, to 
generate a foliation-preserving diffeomorphism, requires
\begin{equation}
    \epsilon^0_{,i} = 0\ .
        \label{diffeo1}
\end{equation}


\subsection{Expressing the homogeneous metric}
\label{subsec:1step.kill}

Recall the commutation relations of the Killing vectors equation 
(\ref{1.com}).  At this point it is customary to introduce a set of 
three independent, right-invariant (under the Lie algebra) vector 
fields ${\bf Y}_a = Y_a^i(t,x)\bbox{\partial}_{i}$, which satisfy
\begin{equation}
    \left[{\bf K}_a,{\bf Y}_b\right] = 0\ .
        \label{invarbasis}
\end{equation}
We take them to be tangent to the homogeneous hypersurfaces.  They 
also define a Lie algebra and can be taken such that
\begin{equation}
    [{\bf Y}_{a},{\bf Y}_{b}] = - C^{c}_{ab}{\bf Y}_{c}\ .
        \label{commbasis}
\end{equation}
These commutation relations are ensured if we take ${\bf Y}_a$ to 
coincide with ${\bf K}_a$ at an arbitrary point in $\Sigma_{t}$ and 
define them at other points in $\Sigma_{t}$ by equation 
(\ref{invarbasis}) \cite{taub51}.

In each $\Sigma_{t}$ we can define a basis of 1-forms, 
$\bbox{\omega}^{a} = \omega^a_i(t,x) {\bf d} x^i$, dual to these 
left-invariant vectors: 
$\bbox{\omega}^{a}({\bf Y}_{b}) = \delta^{a}_{b}$.  The Lie algebra 
property becomes (${\bf d}_3$ is differentiation with respect to the 
space variables)
\begin{equation}
    {\bf d}_{3}\bbox{\omega}^{a} 
    = {1\over2} C^{a}_{bc}\bbox
        {\omega}^{b} \wedge \bbox{\omega}^{c}\ .
            \label{curlforms}
\end{equation}

The metric can be written, using the anholonomic basis 
$\{{\bf d}t,\bbox{\omega}^{a}\}$, as
\begin{equation}
    {\bf g} = - N^2 {\bf d}t^2 
        + g_{ab}\left(N^a {\bf d} t + \bbox{\omega}^{a}\right)
            \left(N^b {\bf d} t + \bbox{\omega}^{b}\right)\ ,
                \label{lapseshift}
\end{equation}
where $N$ is the lapse function and $N^b$ the shift 
vector variables in this basis. 
The Killing conditions on $\bf g$ give 
($~\dot{}$ denotes $\partial/\partial t$)
\begin{mathletters}\label{killmetric}
\begin{equation}
    {\bf K}_a(g_{bc}) = 0 \Longleftrightarrow g_{ab} = g_{ab}(t)\ ,
        \label{killmetric.a}
\end{equation}
\begin{equation}
    {\bf K}_a(N) = 0 \Longleftrightarrow N = N(t)\ ,
        \label{killmetric.b}
\end{equation}
\begin{eqnarray}
	{\cal L}_{{\bf K}_a}(N^b {\bf d}t & + & \bbox{\omega}^b) = 0 
	\nonumber \\
	\Longleftrightarrow \dot{\bf K}_a & = & 
	    \left[N^b {\bf Y}_b, \, {\bf K}_a\right] 
	    = -{\bf K}_a(N^b){\bf Y}_b\ .
	\label{killmetric.c}
\end{eqnarray}
\end{mathletters}%
The first two results are clear: no spatial dependence for $N$ and 
$g_{ab}$.  The third result links the time dependence of the Killing 
vectors to the spatial dependence of the shift variables $N^a$.  It 
relates dimensional reduction and gauge-fixing reduction.  Since the 
shift variables are completely arbitrary because of the gauge freedom, 
namely diffeomorphisms satisfying $\epsilon^0_{,i} = 0$ (see equation 
(\ref{diffeo1})), we may consider this third relation as giving the 
time evolution of the Killing vector components for a given set of the 
shift variables.  This shows in particular that if for an ``initial'' 
time, say $t=0$, the Killing conditions are satisfied, they will be 
satisfied in future times, the third relation providing us with the 
form which the Killing vectors take.

At this point we could proceed to reduce the Lagrangian (which we will 
describe below).  But reduction within the Lagrangian and reduction in 
the equations of motion are procedures that in general do not commute, 
as we could immediately verify.  In the next section a thorough study 
of this non-commutativity will be given.

The fact that the shift variables depend arbitrarily on the space 
coordinates shows that we must further reduce the gauge freedom in 
order to end up with the reduced gauge group of 
time-reparameterization invariance.  To do so, the simplest way is to 
require the shift variables to be spatially constant.  According to 
the third relation above, this is equivalent to requiring time 
independence of the Killing vectors.  This is our second step in the 
gauge fixing procedure, in the next section, and is a case of 
adapting spatial coordinates to the Killing structure.


\section{Space}
\label{sec:2step}

We introduce the partial gauge-fixing conditions 
\begin{equation} 
    N^a_{,i} = 0\ .
        \label{n(t)}
\end{equation}
We know from the previous analysis, equation (\ref{killmetric.c}), 
that the Killing vectors become time independent,
\[ 
    {\bf K}_a = K_a^i(x) \bbox{\partial}_i\ , 
\]
and we can also choose the right invariant vectors ${\bf Y}_a$ and the 
1-forms $\bbox{\omega}^a$ to be so.  It was realized in 
\cite{jantzen79} that the shift vector 
${\bf N} \equiv N^a(t){\bf Y}_a$ is a one-parameter family of inner 
automorphisms of the right-invariant Lie algebra (generators 
${\bf Y}_b$).  More details are given in Section \ref{sec:4step}.
Now the metric is written as
\begin{eqnarray}
    {\bf g} &=& - N^2(t) {\bf d}t^2 \nonumber \\
      &~& + g_{ab}(t) \left(N^a(t){\bf d}t + \bbox{\omega}^a\right)
           \left(N^b(t){\bf d}t + \bbox{\omega}^b\right)\ ,
               \label{tmetric}
\end{eqnarray}
with
\[
    \bbox{d\omega} 
    = {1\over2}C^a_{bc}\bbox{\omega}^b\wedge\bbox{\omega}^c\ ,
\]
and the Killing conditions are built in. 

We proceed to study the non-commutativity of the operations of 
introducing the conditions $g_{ab,i} = N_{,i} = N^a_{,i} = 0$ into the 
equations of motion (that is, to look for a restricted set of 
solutions of the Einstein equations) or directly into the Lagrangian.  
The calculation given in Appendix \ref{app:homog} shows explicit 
examples of the ideas we discuss in this Section.  In the 
anholonomic basis $\{{\bf d}t,\,\bbox{\omega}^a\}$, the 
Einstein-Hilbert Lagrangian density, once a total divergence is 
discarded, can be written as (our notations follow \cite{wald84})
\begin{equation}
    {\cal L} = |\omega|\sqrt{g} N ({}^3\!R + K_{ab}K^{ab} - K^2) 
    \equiv |\omega|\tilde{\cal L} \ ,
        \label{lagr}
\end{equation}
where $|\omega|=\det(\omega^a_i)$, $g=\det(g_{ab})$, $N \neq 0$ is the 
lapse function, ${}^3\!R$ is the three-metric curvature, $K_{ab}$ the 
extrinsic curvature
\begin{equation}
    K_{ab}={1\over 2N}(\dot g_{ab} - N_{a|b} - N_{b|a}) \ ,
    \label{extcurv}
\end{equation} 
and $K=e^{ab}K_{ab}$ its trace (${}_{|}$ denotes the spatial covariant 
derivative and $e^{ab}$ is the matrix inverse of $g_{ab}$).  Notice 
that ${\cal L}$ is second order in the three-space derivatives.

One could also add other fields in the Lagrangian and implement the 
Killing conditions on them.  For instance, in the Einstein-Maxwell 
case, the one-form gauge field 
${\bf A} = A_0 {\bf d}t + A_a \bbox{\omega}^a$ must satisfy 
$A_{0,i} = A_{a,i} = 0$ (see Appendix \ref{app:maxwell}).  If we label 
as $X$ the generic variable $g_{ab}, N, N^a, A_0, A_a, ...$, then 
$\tilde{\cal L}$ is a function 
$\tilde{\cal L}(X, \dot X, {\bf Y}_bX, {\bf Y}_a{\bf Y}_bX )$, 
where the last argument in $\tilde{\cal L}$ has $a \geq b$.  Notice 
that ${\bf Y}_a{\bf Y}_bX$ contains not only second order but also 
first order spatial derivatives of $X$.

Let us write the Euler-Lagrange equations using 
the variables just displayed. The variation $\delta{\cal L}$ is
\begin{eqnarray*}
	\delta{\cal L} & = & {\partial{\cal L}\over\partial X}\delta X
	    + {\partial{\cal L}\over\partial\dot X}\delta\dot X
	    + {\partial{\cal L}\over\partial{\bf Y}_aX}  
	        \delta ({\bf Y}_aX)  \\
	 & ~ & + \sum_{a \geq b}
	     {\partial{\cal L}\over\partial{\bf Y}_a{\bf Y}_bX}
            \delta ({\bf Y}_a{\bf Y}_bX)\ .
\end{eqnarray*}
Integration by parts, dropping the boundary terms, and repeated use 
of the relation
\begin{equation}
    {\bf Y}_a (|\omega|) + |\omega|\bbox{\partial}_j Y_a^j 
    = |\omega| C^d_{ad}\ , 
        \label{omega}
\end{equation}
yields the following form for the Euler-Lagrange equations: 
\begin{eqnarray}
	{\delta{\cal L}\over\delta X} & = & |\omega|
	    \bigg({\partial\tilde{\cal L}\over\partial X}
	       -\partial_{t}{\partial\tilde{\cal L}\over\partial\dot X}
	       -({\bf Y}_a + C_{ac}^c)
	           ({\partial\tilde{\cal L}\over\partial{\bf Y}_aX}) 
	\nonumber  \\
	 & ~ & + \sum_{a \geq b} 
	     ({\bf Y}_b+C_{bd}^d)({\bf Y}_a + C_{ac}^c)
	      ({\partial\tilde{\cal L}\over\partial{\bf Y}_a{\bf Y}_bX}) 
	     \bigg). 
	\label{e-l}
\end{eqnarray}
Due to the particular structure of equation (\ref{lagr}), the third 
and fourth order spacetime derivatives in equation (\ref{e-l}) will 
cancel, but this fact does not alter our discussion.

If we define the reduced Lagrangian
\[
    {\cal L}_R(X, \dot X) 
    \equiv \tilde{\cal L}(X, \dot X, 
        \,{\bf Y}_aX =0,\, {\bf Y}_a{\bf Y}_bX = 0)\ , 
\]
we end up with
\begin{eqnarray}
	\big({\delta{\cal L}\over\delta X}\!\!\! & \big) &
        \!\!\!_{_{(\partial_i X = \partial_{ij}X = 0)}} \nonumber\\
    & = & |\omega|\bigg({\delta{\cal L}_{R}\over\delta X}
        -C_{ac}^c
          \big({\partial\tilde{\cal L}\over\partial{\bf Y}_aX}\big)
            _{_{(\partial_i X = \partial_{ij}X = 0)}}
	\nonumber  \\
	 & ~ & \quad
	      + \sum_{a\geq b} C_{ac}^c C_{bd}^d
        \big({\partial\tilde{\cal L}\over
            \partial{\bf Y}_a{\bf Y}_bX} \big)
            _{_{(\partial_i X = \partial_{ij}X = 0)}} 
    \bigg)\ .
	\label{noncom}
\end{eqnarray}
equation (\ref{noncom}) displays the non-commutativity between the two 
procedures: Reduction of the Lagrangian through the Killing conditions 
or reduction of the equations of motion.  As we have mentioned before, 
this problem was identified long ago.  The distinction between Class A 
and Class B Bianchi models is crucial in this respect, for Class A is 
characterized by the vanishing of the trace $C_{ab}^b$ 
\cite{ellis/maccalum69}.

Summing up, we see that the implementation of the Killing conditions 
within the Lagrangian produces no harm, that is, commutativity holds 
in equation (\ref{noncom}), for Class A models.  For Class B models 
the equations of motion derived from ${\cal L}_R$ are wrong when there 
is a contribution different from zero coming from the last two pieces 
in equation (\ref{noncom}).  One may wonder whether this means that 
the situation is hopeless if we want to have a Lagrangian formulation 
for Class B models.  Perhaps in general, but in some special cases 
there may be {\it ad hoc} solutions.  In Appendix \ref{app:maxwell} we 
show an example of such a case within homogeneous Maxwell theory.


\subsection{Gauge transformations after the second step}
\label{subsec:2step.gauge}

At this stage we have time-independent Killing vectors.  Let 
${\bf E} = \epsilon^{\mu}\bbox{\partial}_{\mu} 
    = \epsilon^{0}\bbox{\partial}_{0} 
        + \epsilon^{a}Y^{i}_{a}\bbox{\partial}_{i}$ 
be a generator of a diffeomorphism which preserves this requirement 
as well as equation (\ref{diffeo1}):
\begin{equation}
	[{\bf E},{\bf K}_{a}]\dot{} = [\dot{\bf E},{\bf K}_{a}] = 0\ ,\ 
	\epsilon^{0}=\epsilon^{0}(t)\ ,
	                                        \label{notimeepsil}
\end{equation}
where $\dot{}=\partial/\partial t$.  The general form for 
the spatial components of {\bf E} is given by 
\begin{equation}
	\epsilon^a(t,x) = \varphi^a(t) + \zeta^a(x)\ ,
	                                        \label{diffeo2}
\end{equation}
with $\varphi^a(t)$ and $\zeta^i(x)$ arbitrary functions of time and 
spatial coordinates respectively. 

Suppose ${\bf E}_{1}$ and ${\bf E}_{1}$ are two generators 
satisfying equation (\ref{notimeepsil}):
\[   [\dot{\bf E}_{1},{\bf K}_{a}] 
            = [\dot{\bf E}_{2},{\bf K}_{a}] = 0\ .
\]
By the Jacobi identity it is clear that 
$\bigr[[\dot{\bf E}_{1},\dot{\bf E}_{2}],{\bf K}_{a}\bigr] = 0$,
but it is not necessarily the case that 
$\bigr[[{\bf E}_{1},{\bf E}_{2}]\,\dot{},{\bf K}_{a}\bigr] = 0$.
A closure process which is more general than commutation applies for 
these generators.  The infinitesimal diffeomorphism generated by 
${\bf E}_{1}$ changes the Killing vector fields (in the active view of 
diffeomorphisms), but it also changes the invariant basis vector 
fields and the second generator ${\bf E}_{2}$ as well.  These changes, 
particularly changes in the invariant basis vectors ${\bf Y}_{a}$, 
mean that the form of equation (\ref{diffeo2}) cannot be expected to 
be invariant during the process of commutation.  We will not pursue 
this matter further, for in the next step, where the shift vector is 
set to zero, the form of the gauge group once again becomes 
straightforward.

Notice that for Class A models, the first term on the right in 
equation (\ref{diffeo2}) is a Noether symmetry for ${\cal L}_R$.  Our 
Lagrangian ${\cal L}$, equation (\ref{lagr}), differs from the 
Einstein-Hilbert Lagrangian ${\cal L}_{EH}$ by a divergence
\[
    {\cal L} = {\cal L}_{EH} +|\omega|\sqrt{g} \, m^a\!_{|a}\ .
\]
General reparameterization invariance under a diffeomorphism 
generated by 
${\bf v}=\epsilon^\mu(x)\bbox{\partial}_\mu=\varphi^a(t){\bf Y}_a$ 
produces the functional variation of ${\cal L}_{EH}$,
\[
    \delta_{\bf v} {\cal L}_{EH} 
    = \bbox{\partial}_\mu (\epsilon^\mu {\cal L}_{EH})\ .
\]
Therefore,
\begin{equation}
    \delta_{\bf v} {\cal L} 
    = \bbox{\partial}_\mu (\epsilon^\mu {\cal L}_{EH}) 
      + \delta_{\bf v} \left(|\omega|\sqrt{g} 
          ({\bf Y}_a m^a + C^b_{ab} m^a) \right)\ .
                \label{rep-inv}
\end{equation}

Let us now reduce these expressions, for Class A models, introducing 
our second step gauge fixing, 
$\bbox{\partial}_i X = \bbox{\partial}_{ij}X = 0$, where 
$X = g_{ab}, N, N^a, A_0, A_a, \ldots\,$.  The left side of equation 
(\ref{rep-inv}) becomes
\[
    |\omega| \delta_{\bf v} {\cal L}_R\ .
\]
The first term on the right in equation (\ref{rep-inv}) is 
\begin{eqnarray*}
    &~&\left(\bbox{\partial}_\mu(\epsilon^\mu{\cal L}) \right)
        _{_{(\partial_i X = \partial_{ij}X = 0)}} \\
    &~&\qquad = f^a(t) \bbox{\partial}_i 
      \left( |\omega| Y_a^i \tilde{\cal L} \right)
          _{_{(\partial_i X = \partial_{ij}X = 0)}} = 0\ , 
\end{eqnarray*}
where we have used equation (\ref{omega}) in the last step.  The 
second term on the right in equation (\ref {rep-inv}) becomes
\begin{eqnarray*}
    &~& \left(\delta_{\bf v}\left(|\omega|\sqrt{g}\, 
        m^a\!_{|a} \right) \right)
            _{_{(\partial_i X = \partial_{ij}X = 0)}}  \\
    &~&\qquad = |\omega| {\bf Y}_a \left(\delta_{\bf v} 
        \left(\sqrt{g}\, m^a \right) \right)
            _{_{(\partial_i X = \partial_{ij}X = 0)}} = 0 \ .
\end{eqnarray*}
We end up with $\delta_{\bf v} {\cal L}_R = 0$; this is a gauge 
Noether symmetry provided by the reduced theory.  

The second term on the right in equation (\ref{diffeo2}) is not
retrievable as a gauge symmetry from the reduced Lagrangian
formulation (except when it is a copy, with constants $\varphi^a$,
of the first term).  However, we must still provide a gauge fixing for
it.  In step four of our gauge fixing procedure we will deal with it.
It is obvious from equation (\ref{diffeo2}) that the process 
of reducing the gauge group is not yet finished; we now continue 
this task.


\section{Zero shift}
\label{sec:3step}

The remaining diffeomorphism invariance allows us to perform a new 
partial gauge-fixing by introducing a set of constraints that helps to 
eliminate some of the arbitrariness that exists in our equations of 
motion.  To simplify the analysis we consider the case in which the 
metric is the only field in ${\cal L}$.  Should other gauge fields be 
present, the dynamics could have additional arbitrariness.

Notice that the Lagrangian is independent of $\dot N$ and $\dot N^a$.  
This fact implies that the dynamical evolution vector field operator 
in configuration-velocity space has \cite{batlle/al86,pons/shepley95} 
a term of the type (see Appendix \ref{app:varprin} for clarification 
of this point)
\begin{equation}
    \int d^3 x \left( \eta^0(t,x) {\partial\over\partial \dot N} 
       + \eta^a(t,x) {\partial\over\partial \dot N^a} \right) \ ,
           \label{freed}
\end{equation}
with $\eta^0, \eta^a$ arbitrary functions.  The important consequence 
of this arbitrariness is that to fix the dynamics we must fix the 
values of $N$ and $N^a$.  Here we are only interested in a partial 
gauge-fixing: We only fix the shift functions.  The simplest way of 
doing so is by introducing the gauge-fixing constraints $N^a = 0$.  It 
is always possible to pass from an initial configuration with 
$N^a\neq0$ to a final configuration with $N^a = 0$ by using 
diffeomorphisms satisfying the restrictions of equation 
(\ref{diffeo2}); the geometric picture of this transformation is to 
make the curves in ${\cal M}$ tangent to the normal vector of the 
$\Sigma_{t}$ coincident with the curves generated by 
$\bbox{\partial}_t$.

This is the Lagrangian version of the partial gauge fixing.  Stability 
of the new constraints under time evolution yields the new constraints 
$\dot N^a = 0$; then requiring stability again will make the arbitrary 
functions $\eta^a = 0$.

To complete the picture, let us go back to the second gauge fixing 
step, equation (\ref{n(t)}).  Consider equation (\ref{freed}) again.  
Stabilization of $N^a_{,i} = 0$ implies ${\dot N}^a_{,i} = 0$.  The 
stabilization of ${\dot N}^a_{,i} = 0$ yields $\eta^a_{,j} = 0$.  By 
the same token, $\eta^0_{,j}$ vanishes as a consequence of the 
relation, which is also a gauge fixing, $N_{,i} = 0$, obtained in the 
first gauge fixing step.  Therefore, from the point of view of the 
reduction of the equations of motion, the arbitrariness in the 
dynamical evolution vector field operator is described by four 
functions $\eta^0 (t), \, \eta^a (t)$.  The gauge freedom associated 
with this arbitrariness is given by $\epsilon^0(t), \varphi^a(t)$ in 
equation (\ref{diffeo1}) and equation (\ref{diffeo2}).  Notice that 
the second term in equation (\ref{diffeo2}) is unretrievable from the 
reduced dynamics.  This is another way to verify the limitations of 
the reduced formalism under the Killing conditions already pointed 
out in the last paragraphs of Section \ref{sec:2step}.


\subsection{Gauge group after the third step}
\label{subsec:3step.gauge}

Once the shift functions $N^a$ have been set to zero, the gauge group 
has been greatly reduced, for the only remaining diffeomorphisms 
still available to us are those that, besides keeping the 
three-foliation, preserve the conditions $N^a = 0$.  This requirement 
results in
\begin{equation}
    0 = \dot\epsilon^a\ .
        \label{diffeo3}
\end{equation}
The remaining diffeomorphisms therefore are such that $\epsilon^0$ 
only depends on $t$ (see equation \ref{diffeo1}), whereas the 
$\epsilon^a$ depend only on the three-space coordinates $x^{i}$ (see 
equation \ref{diffeo3}).  A nice picture emerges: The remaining gauge 
group has been factorized into two commuting subgroups, the group of 
time reparameterizations and the group of three-space diffeomorphisms.

Properly speaking, only the first group is still a gauge group; it is 
directly associated with the freedom that is left in the Lagrangian 
evolution operator and that is displayed in the arbitrariness of 
$\eta^0$ in equation (\ref{freed}) ($\eta^a$ are zero in our 
particular gauge-fixing).

The group of three-space diffeomorphisms is not a gauge group because 
it does not have room for what is characteristic of the gauge freedom: 
to connect the members of a family of field configurations, all of 
which are solutions of the equations of motion, that share the same 
set of initial-value conditions.  This group must be understood as 
describing a redundancy in the space of initial-conditions for our 
theories.  We call this group a residual gauge group.  A 
parallel case in electromagnetism is the residual gauge symmetry that 
is left after the introduction of the Lorentz gauge 
$\bbox{\partial}_\mu A^\mu = 0$: The transformation $A_\mu\rightarrow 
A_\mu+\partial_\mu\Lambda$ is a residual gauge symmetry if $\Box 
\Lambda = 0$.  In any case, one must take into account that the gauge 
fixing procedure is only finished when we have completely removed 
these residual gauge transformations \cite{pons/shepley95}.  Further 
comments on this important point are made in the last section of this 
paper.

We finish this section by observing that the comments raised in the 
previous section still apply here: Only the diffeomorphisms of the 
form $\epsilon^i = B^a {\bf Y}_a^i$, with $B^a$ constant are 
obtainable as gauge Noether symmetries from the reduced Lagrangian in 
Class A models.  In the language introduced in the next section, these 
vector fields $B^a {\bf Y}_a^i$ define the inner automorphisms for the 
right invariant Lie algebra (generators ${\bf Y}_a$)


\section{Residual three-diffeomorphisms}
\label{sec:4step}

Now we are ready to perform the fourth step of the gauge fixing.  
Notice that the three-metric, 
$g_{ab}(t)\bbox{\omega}^a\bbox{\omega}^b$ on the surfaces $\Sigma_{t}$ 
is also Killing with respect to the vectors ${\bf K}_a$.  We are going 
to fix, in principle, the residual three-diffeomorphism group by 
working from now on with this basis of one-forms 
$\{\bbox{\omega}^a\}$.  In the active view, a general 
three-diffeomorphism will drive ${\bf K}_a \rightarrow {\bf K'}_a$, 
$\bbox{\omega}^a \rightarrow {\bbox{\omega}'}^a$.  The transformed 
metric ${g}_{ab}(t) {\bbox{\omega}'}^a {\bbox{\omega}'}^b \equiv 
{g'}_{ab}(t,x) {\bbox{\omega}}^a {\bbox{\omega}}^b$ 
will be Killing with respect to the new set ${\bf K'}_a$ and will in 
general no longer be Killing with respect to the original vectors 
${\bf K}_a$.  In other words, ${g}'_{ab}$ will in general acquire a 
dependence on the spatial coordinates.  However, there is another 
possibility to explore: the case when the new set ${\bf K'}_a$ happens 
to belong to the Lie algebra generated by the original Killing vectors 
${\bf K}_a$.  This is equivalent to saying that
\begin{equation}
    {\bbox{\omega}'}^a = M^a_{{}b} \bbox{\omega}^b \ ,
        \label{m}
\end{equation}
with $M^a_b$ constant.  These are the homogeneity-preserving 
diffeomorphisms (HPD), which were introduced by Ashtekar and Samuel 
\cite{ashtekar/samuel91}.

Generators ${\bf v}$ for the HPD are better expressed in the basis of 
the invariant vectors ${\bf Y}_a$, \, 
${\bf v} = f^a(x){\bf Y}_a$.  The condition for an HPD is 
(${\cal L}$ is Lie derivative here)
\begin{equation}
    {\cal L}_{\bf v} \bbox{\omega}^b = B^b_{a} \bbox{\omega}^a\ , 
        \label{hpd-y}
\end{equation}
with $B^b_{a}$ a constant matrix. This is equivalent to 
\begin{equation} 
    {\cal L}_{\bf v} {\bf K}_a = -A^b_{a} {\bf K}_b\ , 
        \label{hpd-k}
\end{equation}
with $A^b_{a}$ a constant matrix.  The left-invariant Lie algebra 
(generators ${\bf K}_a$) and the right-invariant Lie algebra 
(generators ${\bf Y}_a$) possess the same automorphisms.  If the point 
$x_0$ in $\Sigma$ is the point where ${\bf Y}_a$ coincides with 
${\bf K}_a$, then $B^b_{{}a} = A^b_{{}a} + f^c(x_0) C^b_{ca}$.

Equation (\ref{hpd-k}) is
\begin{eqnarray}
    &~& ({\bf K}_a f^c){\bf Y}_c 
    = -[{\bf v},{\bf K}_a] = A^b_{{}a} {\bf K}_b \nonumber\\
    &~&\qquad \Longleftrightarrow
      ({\bf K}_a f^c)= ({\bf K}_b\cdot\bbox{\omega}^c) A^b_{{}a}\ .
\end{eqnarray}
This last equation is equivalent to
\begin{equation}
    {\bf d}f^c 
    = ({\bf K}_b \cdot \bbox{\omega}^c) A^b_{{}a} {\bf K}^a\ , 
        \label{hpd-eq}
\end{equation}
where ${\bf K}^a$ are the dual forms to ${\bf K}_b$.  Notice that 
$({\bf K}_b \cdot\bbox{\omega}^c)$ is the adjoint representation of 
the Lie group expressed in the local patch we are working in.  Because 
${\bf dK}^c = -{1\over2}C_{ab}^c {\bf K}^a \wedge {\bf K}^b$ and 
${\bf d}({\bf K}_d \cdot \bbox{\omega}^c) 
= C_{bd}^e ({\bf K}_e \cdot \bbox{\omega}^c) {\bf K}^b$, 
the local integrability conditions for equation (\ref{hpd-eq}),
\begin{equation}
    {\bf d}\left(({\bf K}_b\cdot\omega^c) A^a_{{}b}{\bf K}^a\right)
    = 0\ , 
        \label{hpd-int}
\end{equation}
read
\begin{equation}
    C_{eb}^a A^e_{c} - C_{ec}^a A^e_{b} + C_{bc}^e A^a_{e} = 0\ , 
        \label{cacaca}
\end{equation}
which are the conditions for Lie algebra automorphisms.  Trivial 
solutions are $A^a_{b}=0$ and $A^a_{b} = A^c C_{cb}^a $, with $A^c$ 
constant.  The solutions for the first case are
\begin{equation}
    {\bf v} = B^c {\bf Y}_c \ ,
        \label{inner-y}
\end{equation}
with $B^c$ constant, and for the second, 
\begin{equation}
    {\bf v} = A^c {\bf K}_c \ .
        \label{inner-k}
\end{equation}
The first (second) solutions describe the inner automorphisms for the 
right-invariant (left-invariant) Lie algebra.  In finite form, the 
$M^a_{{}b}$ in equation (\ref{m}), (${\bf M} = e^{\bf B}$, {\bf B} of 
equation \ref{hpd-y}), satisfy
\begin{equation}
    ({\bf M}^{-1})^d_a ({\bf M}^{-1})^e_b\, C_{de}^f M^c_f 
    = C^c_{ab} \ .
        \label{autom}
\end{equation}

The integrability conditions equation (\ref{cacaca}) only guarantee 
the local existence of the HPD. But we need these HPD to be global 
diffeomorphisms if equation (\ref{autom}) is to hold everywhere.  Both 
the Lie algebra structure and the topology of the surfaces of 
homogeneity play a role in the determination of the existing HPD. In 
particular, for globally homogeneous cosmologies with simply connected 
surfaces of homogeneity, every solution of equation (\ref{cacaca}) 
defines a generator of a HPD. The relevant role of spatial topology 
was first emphasized in \cite{ashtekar/samuel91} and treated in great 
detail in 
\cite{koike/tanimoto/hosoya94,tanimoto/koike/hosoya97a,%
tanimoto/koike/hosoya97b}. 
 The degrees of freedom for Class A globally homogeneous Bianchi 
models are also discussed in this reference; we will return to the 
question of degrees of freedom later.


\subsection{Gauge group after the fourth step}
\label{subsec:4step.gauge}

Once we have completely eliminated the residual three-diffeomorphism 
gauge invariance, the remaining gauge group is that of time 
reparameterizations.  The lapse variable transforms, under the time 
reparameterization gauge group, as a scalar density:
\begin{equation}
    \delta_{\epsilon^0 (t)} N 
    = \epsilon^0 \dot N + {\dot \epsilon^0} N \ , 
        \label{del-n2}
\end{equation}
and the three-metric variables $g_{ab}$ as a scalar:
\begin{equation}
    \delta_{\epsilon^0 (t)} g_{ab} = \epsilon^0 {\dot g}_{ab}\ . 
\end{equation}


\section{Consequences of $N^{\lowercase{a}} = 0$}
\label{sec:zeroshift}

We have seen in Section \ref{sec:2step} the non-commutativity between 
the operations of inserting the Killing conditions into the equations 
of motion or directly into the Lagrangian.  Here we are going to 
consider another kind of non-commutativity, the one coming from 
plugging the gauge fixing $N^a = 0$ into the Lagrangian.  To make 
things clear, we only consider here the reduction of ${\cal L}$ in 
equation (\ref{lagr}) to a reduced Lagrangian ${\cal L}_{\rm GF}$, 
prior to the introduction of the Killing conditions in it (GF stands 
for gauge fixing).

The important question is whether this new Lagrangian 
${\cal L}_{\rm GF}$ is going to reproduce the same equations of motion 
derived from ${\cal L}$ under the gauge-fixing conditions $N^a = 0$.  
The answer, from a general perspective which is summarized in Appendix 
\ref{app:varprin}, can be found in \cite{pons96}, and in general is in 
the negative.

Bringing the results of \cite{pons96} to our case, we find 
(here we use $[{\cal L}]$ for the Euler-Lagrange variation): 
\begin{equation}
    [{\cal L}] = 0 \ , \ N^a = 0
    \Longleftrightarrow
    [{\cal L}_{\rm GF}] = 0 \ , \ {\cal H}_a = 0 \ , 
        \label{equiv}
\end{equation}
where ${\cal H}_a$ are the Hamiltonian momentum constraints in the 
anholonomic basis and expressed in coordinate-velocity space. The 
$N^a$ have been set to zero in ${\cal H}_a$. If we also implement 
the Killing conditions in ${\cal H}_a$, we get 
\begin{eqnarray}
    {\cal H}_a 
    &=& - |\omega| {\sqrt{g}\over N} 
      (\dot g_{bc|a} - \dot g_{ac|b}) g^{bc} 
         \nonumber \\ 
    &=& |\omega| {\sqrt{g}\over N} 
        (C^c_{ab} k^b_c + C^b_{bc}k^c_a) \ ,
        \label{mom-a}
\end{eqnarray}
where $k^b_c = g^{bd}\dot g_{dc}$.

Therefore, if we use the completely reduced Lagrangian, with gauge 
fixing plus Killing conditions, with configuration variables 
$N, g_{ab}$, we must be aware that the correct equations of motion 
will require that initial conditions be taken such that
\begin{equation}
    C^c_{ab} k^b_c + C^b_{bc}k^c_a = 0\ .
        \label{h=0}
\end{equation}
We show in Appendix \ref{app:varprin} that if the initial conditions 
satisfy equation (\ref{h=0}), then equation (\ref{h=0}) will hold at 
any time, provided the equations of motion are satisfied.

In Class A models, where $C^b_{bc}= 0$, equation (\ref{h=0}) has only 
the term $C^c_{ab} k^b_c = 0$.  It is worth noting that, 
except for Type I models, which have vanishing structure constants, in 
all other Class A models, equation (\ref{h=0}) must be enforced on the 
initial conditions.  A diagonal form for the metric, for instance, 
will guarantee the fulfillment of equation (\ref{h=0}) for most Class 
A models as long as the structure constants are taken in the form 
displayed in \cite{ryan/shepley75} (the exception is the group of Type 
VI${}_{-1}$).  However, this diagonal form may not exhaust the 
possible physics available in these models, and we do not assume 
diagonality here.

Notice that the HPD cannot be used to make the initial conditions
satisfy equation (\ref{h=0}):  Under a change 
\[ 
    g_{ab} \rightarrow {g'}_{ab} = M_a^c g_{cd}M^d_b\ ,
\]
the momentum constraints change as
\[ 
    {\cal H}_a({\bf g}') = (\det{\bf M}) M^b_a {\cal H}_b({\bf g}) \ .
\]

The result of our analysis is the following: The completely reduced 
Lagrangian gives the correct equations of motion for Class A models if 
and only if we tune the initial conditions (at $t=0$) $g_{ab}(0)$, 
${\dot g}_{ab}(0)$ in such a way that
\begin{equation}
    C^c_{ab} g^{bd}(0) \dot g_{dc}(0) = 0 \ .
        \label{h=0-0}
\end{equation}

To this result we must add the fact that two three-metrics, $g_{ab}$ 
and ${g'}_{ab}$, are physically equivalent, that is, gauge related, if 
there is an HPD such that in the notation of equation (\ref{m}) 
${g'}_{ab}=M_a^c g_{cd}M^d_b$.  We must implement this fact in 
counting the number of independent initial conditions.  In order to 
have a correct dynamics and a correct counting of the degrees of 
freedom for Class A Bianchi models we must require that (1) the 
initial conditions must be chosen such that equation (\ref{h=0-0}) is 
satisfied, and (2) initial conditions related by HPD must be 
considered physically equivalent.


\section{Hamiltonian approach}
\label{sec:hamiltonian}

The standard Hamiltonian ADM approach \cite{ryan/shepley75} is built 
upon the Lagrangian equation (\ref{lagr}).  Since ${\cal L}$ does not 
depend on $\dot N$ and ${\dot N}^a$, the conjugate variables 
${\cal P}_0$ and ${\cal P}_a$ to $N$ and $N^{a}$ are the primary 
constraints in phase space.  The canonical Hamiltonian has the form
\begin{equation}
    H = \int d^3x \, ( N {\cal H}_0 + N^a {\cal H}_a ) \ .
        \label{theham}
\end{equation}
The Dirac Hamiltonian $H_{\rm D}$ is formed by adding to $H$ a linear 
combination (with arbitrary functions $\lambda_0, \lambda_a $) of the 
primary constraints:
\begin{equation}
    H_{\rm D} 
    = H + \int d^3x (\lambda^0{\cal P}_0 + \lambda^a{\cal P}_a )\ . 
    \label{dirach}
\end{equation}
Stability of the primary constraints under the evolution generated by 
$H_{\rm D}$ leads to constraints which are now secondary:
\begin{equation}
    {\cal H}_0 = 0\ , \ {\cal H}_a =0\ ,
\end{equation}
and no more constraints arise. 

All eight constraints are first class, and the four independent gauge 
transformations that account for the diffeomorphism invariance are 
made out of them.  For the sake of completeness let us write the gauge 
generators which act through the Poisson bracket in the whole phase 
space \cite{pons/salisbury/shepley96} (three-space integrations are 
understood for all repeated indices):
\begin{equation} 
    G(t) = {\cal P}_\mu \dot\xi^\mu 
        + ({\cal H}_\mu 
            + N^\rho D^\nu_{\mu\rho}{\cal P}_\nu) \xi^\mu \ , 
                \label{thegen}
\end{equation}
where the functions $D^\nu_{\mu\rho}$ describe the first class 
structure of the secondary constraints,
\begin{equation}
    \{{\cal H}_\mu,\, {\cal H}_\rho \} 
    = D^\nu_{\mu\rho}{\cal H}_\nu\ , 
\end{equation}
and $\xi^\mu$ are arbitrary functions.  The relationship of the 
transformations generated by $G(t)$ to the standard diffeomorphisms 
generated by a vector field $\epsilon^\mu\bbox{\partial}_\mu$ is given 
by (see \cite{pons/salisbury/shepley96} for instance):
\begin{equation}
    \epsilon^0={\xi^0\over N}\ , \ 
    \epsilon^a=\xi^a - {N^a\over N} \xi^0. 
        \label{project}
\end{equation}

On a given metric, every infinitesimal spacetime diffeomorphism 
$\epsilon^\mu$ is matched by $G(t)$ with the specific functions 
$\xi^\mu$ dictated by equation (\ref{project}).  This means that we 
have exactly the same gauge group either in the Lagrangian or in the 
Hamiltonian formalism, here extended to the phase space, and that in 
this last formalism the same steps must be taken to reduce the gauge 
group to the time-reparameterization invariance.

The implementation of the Killing conditions and the gauge fixing 
$N^a=0$ into the Hamiltonian equation (\ref{dirach}) leads to the same 
conclusions we have arrived at in the Lagrangian formulation.  In 
particular, only for Class A models is the implementation of the 
Killing conditions into the Hamiltonian correct, in the sense that it 
gives the correct equations of motion.  The reason is that the 
functional derivatives in the Hamiltonian formalism develop terms like 
the ones in equation (\ref{e-l}).  In fact, any of the ${\cal H}_\mu$ 
is of the form
\[
    {\cal H}_\mu = |\omega|\tilde{\cal H}_\mu\ ,
\]
in such a way that the three-space functional derivatives, 
corresponding to the Hamiltonian equations of motion, become (here 
X stands for $g_{ab},\tilde\pi^{ab}$, where $\tilde\pi^{ab}$ is 
the canonical conjugate of $g_{ab}$ in the reduced formalism: 
$\pi^{ab}=|\omega|\tilde\pi^{ab}$)
\begin{eqnarray}
    {\delta{\cal H}_\mu\over\delta X} 
    &=& |\omega| \bigg( {\partial\tilde{\cal H}_\mu\over\partial X} 
        - ({\bf Y}_a + C_{ac}^c)
          ({\partial\tilde{\cal H}_\mu\over\partial{\bf Y}_aX}) 
              \nonumber \\
    &+& \sum_{a\geq b} 
        ({\bf Y}_b + C_{bd}^d)({\bf Y}_a + C_{ac}^c)
          ({\partial^{2}\tilde{\cal H}_\mu\over
              \partial{\bf Y}_a{\bf Y}_bX}) \bigg)\ . 
\end{eqnarray}
(As a matter of fact, only ${\cal H}_0$ depends on second order space 
derivatives.)

To illustrate this last result, let us write the Poisson brackets for 
the momenta ${\cal H}_a$.  They satisfy
\begin{eqnarray}
    &~&\left\{\int d^3x{\cal H}_a \xi_1^a,
            \int d^3x \,{\cal H}_b \xi_2^b \right\}
                \nonumber \\
    &~&\qquad =\int d^3x{\cal H}_c 
        \left(\xi_1^a \xi^c_{2,a} - \xi_2^a \xi^c_{1,a} 
            + C^c_{ab}\xi_1^a\xi_2^b \right).
                \label{mom-const}
\end{eqnarray}
When we implement the Killing conditions on the quantity 
$\int d^{3}x{\cal H}_{a}\xi^{a}$, in order to get rid of the 
three-space dependence, it seems that we are bound to take the 
functions $\xi^a$ as constants (but this is not exactly true, as we 
will see shortly).  This would generate diffeomorphisms of the type 
equation (\ref{inner-y}), namely inner automorphisms for the 
left-invariant Lie algebra.  Then $\tilde{\cal H}_a$ reduces to
\begin{equation}
    {\cal H}^{\rm red}_a = C^c_{ab}\tilde\pi^{bd} g_{dc} 
        + C^b_{bc}\tilde\pi^{cd}g_{da}\ , 
            \label{red-const}
\end{equation}
(this is the phase space version of equation (\ref{mom-a})).  

To  continue, we must first introduce the Poisson brackets for the 
reduced theory.  The reduced Poisson brackets $\{-,-\}_{_{R}}$ are 
defined through a renormalization of the old Poisson brackets 
$\{-,-\}_{_{R}} = V \{-,-\}$, where $V$ is the (perhaps infinite) 
right invariant volume element 
$V = \int \bbox{\omega}^1\wedge\bbox{\omega}^2\wedge\bbox{\omega}^3 = 
\int d^3x |\omega|$.  
Notice that as long as the Killing conditions hold we can invert the 
relation $\pi^{ab} = |\omega|\tilde\pi^{ab}$ to 
$\tilde\pi^{ab} = {1\over V} \int d^3x\,\pi^{ab}$.  The commutation 
relations for ${\cal H}^{\rm red}_a$ are
\begin{eqnarray}
    &&\{{\cal H}^{\rm red}_a,{\cal H}^{\rm red}_b \}_{_{R}}
             = C^c_{ab}{\cal H}^{\rm red}_c  \nonumber \\
        &&\quad  + C^d_{dc} (C^e_{ab}\tilde\pi^{cf}g_{fe} 
           + C^f_{fa}\tilde\pi^{ce}g_{eb} 
           - C^f_{fb}\tilde\pi^{ce}g_{ea} )\ .
               \label{red-comm}
\end{eqnarray}
whereas the reduction of the right side of equation (\ref{mom-const}) 
is, after factoring out the three-space volume,
\[
    C^c_{ab}{\cal H}^{\rm red}_c\ .
\]
It is only for Class A models that we get this result in equation 
(\ref{red-comm}): Only for Class A models do the implementation of the 
Killing conditions and the computation of the Poisson brackets 
commute.


\subsection{Homogeneity-preserving diffeomorphisms, revisited}
\label{subsec:HPD} 

We have said, after equation (\ref{mom-const}), that the 
implementation of the Killing conditions on the generator 
$\int d^3x \,{\cal H}_a \xi^a$ does not require the functions $\xi^a$ 
to be constant.  The reason is as follows.  In the 
computations in equation (\ref{mom-const}) we have been dropping 
boundary terms.  This is correct if our functions $\xi^a$ have a 
compact support on $\Sigma_{t}$.  But we know from the Lagrangian 
analysis that some diffeomorphisms that do not vanish at the 
boundaries, for instance the HPD, play an important role when we 
perform the dimensional reduction through the Killing conditions.

Consider the generator of three-space diffeomorphisms for the reduced 
set of variables $g_{ab},\pi^{ab}$,
\begin{equation}
    \int d^3x \,{\cal H}_a \xi^a\ ,
        \label{gen3}
\end{equation}
with
\begin{equation}
    {\cal H}_a = - |\omega|\sqrt{g} 
       \left(|\omega|^{-1}\sqrt{g}\, \pi^{bc} \right)_{|b} g_{ca}\ .
               \label{gennn3}
\end{equation}
In order to get the standard result
\begin{equation}
    \delta g_{ab} 
    = \left\{g_{ab},\int d^3x {\cal H}_a \xi^a \right\} 
    = \xi_{b|a} + \xi_{a|b}  \ ,
        \label{deltage}
\end{equation}
we always perform integration by parts in equation (\ref{gen3}).  
Since the generator equation (\ref{gen3}) acts locally, what we 
actually do in practice is to compute $\delta g_{ab}(t,x)$ using 
functions that are identical to $\xi_a$ in a neighborhood of a given 
point but that vanish at spatial infinity.  If we perform directly the 
integration by parts in equation (\ref{gen3}) we get
\begin{equation}
    G_3 = \int d^3x \,\pi^{bc} \xi_{c|b}\ , 
        \label{gen3ok}
\end{equation}
which differs from equation (\ref{gen3}) at most by a boundary term.  

Boundary terms may or may not exist depending on the topology of 
$\Sigma$.  As long as gauge transformations on local functions are 
considered, equation (\ref{gen3ok}) is as good as equation 
(\ref{gen3}) to generate the transformations equation (\ref{deltage}), 
but $G_3$ is better when we must work with arbitrary functions $\xi_a$ 
that are different from zero at the boundaries.  Notice that whereas 
equation (\ref{gen3}) vanishes on a solution of the equations of 
motion, because ${\cal H}_a$ is a constraint, equation (\ref{gen3ok}) 
does not need to do so in the case when the functions $\xi_a$ do not 
vanish at the boundaries.  Thus $G_3$ is not bound to be zero on a 
solution of the equations of motion.  It is not a constraint, except 
when there are no boundaries in $\Sigma$ or for some other some 
particular cases that will be examined below.

The dimensional reductions of equation (\ref{gen3}) and equation 
(\ref{gen3ok}) give different result, because integration by parts 
for the space coordinates does not exist in the reduced formalism.  
Now we apply the Killing conditions to equation (\ref{gen3ok}):
\[
    G_3 
    = \int d^3x \,\pi^{bc} \xi_{c|b} 
    = \int d^3x \, |\omega| \tilde\pi^{bc} 
        ({\bf Y}_b \xi_c - \Gamma^d_{bc}\xi_d)\ ,
\]
where the connection coefficients $\Gamma^d_{bc}$ are 
\[
    \Gamma^d_{bc} 
    = {1\over2} \left(g_{ce}C^e_{bf}g^{fd} 
        + g_{be}C^e_{cf}g^{fd} - C^d_{bc} \right)\ ,
\]
(here $g^{fd}=e^{fd}$ since $N^{a}=0$).  Therefore,
\begin{equation}
    G_3 = \int d^3x |\omega| \tilde\pi^{bc} g_{cd} 
        ({\bf Y}_b \xi^d - C^d_{be}\xi^e)\ . 
\end{equation}
Dimensional reduction requires us to factor out the volume 
$V=\int d^3x  |\omega|$.  This means that we need
the remaining piece in the integrand,
$\tilde\pi^{bc} g_{cd} ({\bf Y}_b \xi^d - C^d_{be}\xi^e)$,
to be independent of the space coordinates.  Since $\tilde\pi^{bc}$ 
and $g_{cd}$ are already space independent because they satisfy the 
Killing conditions, we end up with the requirement
\begin{equation}
    {\bf Y}_b \xi^d - C^d_{be}\xi^e = B^d_b\ , 
    \label{7.16}
\end{equation}
with $B^d_b$ a constant matrix.  This condition is exactly equation 
(\ref{hpd-y}), the condition for HPD, expressed in the dual basis.  
Thus we see that the dimensional reduction in phase space has room for 
and only for the HPD we already found in the Lagrangian formalism.  
Then $G_3$ becomes $ G_3 = V B^d_b \tilde\pi^{bc} g_{cd}$, and since 
the reduced Poisson brackets are $\{-,-\}_{_{R}} = V \{-,-\}$, we get 
for the canonical generator of HPD in the reduced formalism,
\begin{equation}
    \tilde G_3 = B^d_b \tilde\pi^{bc} g_{cd}\ .
\label{genhpd}
\end{equation}

We find here another special feature of the Bianchi models, for the 
$\tilde G_3$ in equation (\ref{genhpd}) generates gauge 
transformations, HPD, and yet it is not a constraint of the reduced 
formalism by itself, except for the case in Class A models of inner 
automorphisms.  In this last case, we have $B^d_b = B^c C^d_{cb}$, and 
$\tilde G_3$ for Class A models becomes $\tilde G_3 = B^a {\cal 
H}^{\rm red}_a$, with ${\cal H}^{\rm red}_a$ given in equation 
(\ref{red-const}) with $C^b_{bc} = 0$.  These results completely match 
the ones previously obtained in the Lagrangian analysis.

The generators of HPD that define automorphisms of the Lie algebra 
that are not inner automorphisms (called ``outer'' HPD in 
\cite{ashtekar/samuel91}) are always constants of motion 
\cite{ashtekar/samuel91} of the reduced formalism.  They are enforced 
to be constraints if $\Sigma$ has no boundaries, although these 
constraints cannot be retrieved as such constraints from the reduced 
formalism by itself.  In any case, either being constants of motion or 
constraints, they always generate gauge transformations in the reduced 
formalism.  These transformations, in either the Lagrangian or the 
Hamiltonian picture, are available as gauge transformations from the 
outset, as a consequence of the rationale of the reduction procedure, 
but they cannot be retrieved as gauge transformations from the reduced 
formalism by itself.

A comment is in order for Class B models.  The generators equation 
(\ref{genhpd}) give the HPD for Class B as well as Class A models.  In 
particular, $C^d_{ab}\tilde\pi^{bc}g_{cd}$ generates the inner 
automorphisms associated, in the configuration space picture, with the 
right-invariant vector ${\bf Y}_a$.  Nevertheless, 
$C^d_{ab}\tilde\pi^{bc}g_{cd}$ is not necessarily a constraint for the 
Class B models because the true momentum constraints are those of 
equation (\ref{red-const}), as one can directly verify from the 
equations of motion.  Oddly enough, in Class B models at least one of 
the generators of the residual gauge transformations associated with 
inner automorphisms is not a constraint.  Also, according to equation 
(\ref{red-comm}), these constraints, two of them, are no longer first 
class.  These results hint that the Lagrangian or Hamiltonian 
formulation for Class B models is not possible in general.

Notice that in the case of three-manifolds $\Sigma$ with no 
boundaries, the possible new constraints $\tilde G_3$ in equation 
(\ref{genhpd}) appear also in the Lagrangian formalism by simply 
writing down the process from equation (\ref{gen3}) to equation 
(\ref{gen3ok}) in configuration-velocity variables.

One could think that the mechanism of integration by parts, as used to 
transform equation (\ref{gen3}) into equation (\ref{gen3ok}), and that 
helped us find the whole set of HPD in phase space, could perhaps be 
used in Class B models to transform the reduced ${\cal H}_0$ into the 
correct generator of time translations.  In fact, some unwanted pieces 
can be eliminated this way, but this does not solve the problem.


\subsection{Degrees of freedom}
\label{subsec:degrees}

Notice also that the tangency of the Hamiltonian evolution operator 
equation (\ref{theham}) to the gauge fixing surface defined by 
$N^a = 0$ implies $\lambda^a = 0$.  On the other hand, once the gauge 
fixing $N^a = 0$ has been introduced into $H_{\rm D}$, the stability 
of the primary constraints ${\cal P}_a = 0$ will no longer produce the 
secondary constraints ${\cal H}_a =0$.  This result parallels that of 
equation (\ref{equiv}).  As a matter of fact, and since they have 
become second class constraints, these three couples of canonical 
variables $N^a,{\cal P}_b$, are readily eliminated from the formalism 
by taking the Dirac bracket, which is nothing but the ordinary Poisson 
bracket for the rest of the variables (the same number of variables, 
$N^a,\, {\dot N}^b$ are eliminated at the same stage in the Lagrangian 
formulation).  The important point is that the momentum constraints 
are gone from the reduced formalism, but yet they must be implemented 
(in case they do not vanish identically) from the outset as 
restrictions imposed on the initial conditions, if we want to have the 
right equations of motion.

It is remarkable also that the generators of HPD in the dimensionally 
reduced theory are not necessarily constraints, except for the case of 
inner automorphisms and Class A models.  We still need to fix all the 
residual gauge freedom corresponding to the three-diffeomorphisms, 
because these three-diffeomorphisms have been with us since the 
beginning.  We are in the same situation as we were in the Lagrangian 
formulation, and therefore the same HPD, exactly the same, appear 
here, as we have just shown.  {\it That is why the degrees of 
freedom in both the Lagrangian and the Hamiltonian formalisms 
coincide.}

In \cite{ashtekar/samuel91} a discrepancy between Lagrangian and 
Hamiltonian degrees of freedom is argued on the basis that the only 
HPD available in Hamiltonian formalism are those yielding inner 
automorphisms.  Our analysis differs from the one in 
\cite{ashtekar/samuel91} in that we show that the residual 
three-diffeomorphism invariance that still needs to be fixed 
corresponds to the HPD that we found earlier in the Lagrangian 
formalism.  And this is something that one knows in advance through 
the process of the reducing the original gauge group.

The claims in \cite{ashtekar/samuel91} are opposed in 
\cite{coussaert/henneaux93}, where it is argued that the degrees of 
freedom for both the Lagrangian and Hamiltonian formalisms are the 
same.  However, they base their claim on rejecting the HPD that are 
not associated with inner automorphisms as residual gauge 
transformations that need to be fixed.  In the next section we show 
that all the HPD are indeed residual gauge transformations that 
describe redundancy in the initial conditions of the system and 
therefore do need to be fixed.


\section{Conclusions}
\label{sec:conclusion}

We have proved that Class A Bianchi models allow for a Lagrangian 
formulation as a mechanical particle-like system (finite number of 
degrees of freedom, in contrast to field theory) as long as the 
setting of the initial conditions is taken in a certain way. 

In our transit toward this result we have proceeded as follows:  We 
have carefully produced a partial gauge fixing in four steps, to 
reduce the initial gauge group generated by the four-diffeomorphism 
invariance to time reparameterization invariance.  As a consequence, 
the troubles that beset Class B Bianchi models are identified as 
obstructions to the commutativity between two processes: implementing 
the Killing conditions either into the equations of motion or directly 
into the Lagrangian, followed by deriving the equations of motion.

We distinguish, as regards the Lagrangian or Hamiltonian 
formulations, two types of problems, which correspond, respectively, 
to the implementation of the Killing conditions into the Lagrangian 
and to the implementation of the gauge fixing that sets the shift 
variables to zero.  As we have just said, the first problem prevents 
the Class B Bianchi models from having a reduced Lagrangian or 
Hamiltonian formulation in general.  The second tells us that the 
initial conditions must be chosen to satisfy some former constraints 
(the momentum constraints) of the original theory, even though they 
are no longer constraints for the reduced one.

We have shown that the Hamiltonian formalism has the same gauge 
freedom that is available in the Lagrangian formulation.  The 
reduction of the gauge group in phase space follows exactly the same 
steps as in configuration-velocity space, and the same features, the 
same problems, the same considerations, and the same results apply as 
well.  The fact that the reduced momenta ${\cal H}_a$ in Class A 
models can only generate HPD that are inner automorphisms of the Lie 
algebra has its counterpart in configuration-velocity space in that 
these HPD are the only ones that can be derived as gauge 
transformations from the reduced Lagrangian on its own.  But that does 
not mean that the HPD associated with the ``outer'' automorphisms, if 
there are any, do not need to be quotiented out.  Rather, all HPD must 
be quotiented out, either in the Hamiltonian or in the Lagrangian 
formalism.

Also, what is and what is not a gauge symmetry and the role of the 
residual gauge symmetries has been analyzed in detail.  We think that 
our considerations throw a definitive light on some issues that have 
not yet been settled in the literature 
\cite{ashtekar/samuel91,coussaert/henneaux93}.

Through our analysis, the gauge group is reduced in several stages, 
and residual gauge symmetries appear.  According to the fact that they 
all proceed from the original gauge group, we claim that these 
residual gauge symmetries must be fixed thoroughly.  We think that 
this is an important point that deserves further elaboration below.


\subsection{Gauge fixing for the residual gauge group}
\label{subsec:conc.residual}

Let us show the need to fix the gauge for the residual gauge group of 
three-space diffeomorphisms that appears at the end of the third step 
of the gauge fixing procedure (Section \ref{sec:3step}).  Consider 
that we have a solution $g_{ij}(x,t), N(t)$ of the Einstein 
equations with $N^i=0$ and with initial conditions at $t=0$.  Now 
consider the action of the infinitesimal diffeomorphism transformation 
defined by the vector field $\epsilon^i\bbox{\partial}_i$,
\[ 
    \epsilon^i(x,t) = f(t) \tilde{\epsilon}^i(x), 
\] 
where $f$ has been chosen to satisfy 
\[ 
    f(0) =0\ ,\ \dot f(0) = 0\ ;\ 
    f(t) = 1\ ,\ \dot f(t) = 0 \ {\rm for} \ t \geq 1 \ .
\]

Since this vector field generates a foliation-preserving 
diffeomorphism, it will define a gauge transformation which is still 
allowed in the formalism after the first-step gauge-fixing has been 
done.  The transformed metric $g'_{ij}(x), {N'}(x), {N'}^i(x)$ is also 
a solution of the Einstein equations.  Notice that both metrics 
$g_{\mu\nu}$ and ${g'}_{\mu\nu}$ share the same initial conditions at 
$t=0$.  It is then obvious that the correct interpretation of a gauge 
transformation dictates that $g_{\mu\nu}$ and ${g'}_{\mu\nu}$ are 
physically equivalent and must be so for any other time we take 
for the setting of the initial conditions.

Now consider the relation between these two metrics but with initial 
conditions taken at $t=1$.  They are related by an infinitesimal 
three-space diffeomorphism generated by the time-independent vector 
field $\tilde{\epsilon}^i(x)\bbox{\partial}_i$.  The two metrics 
satisfy the gauge $N^i=0$.  The gauge transformation that connects 
them is not localizable in time (being different from the identity 
only within a finite time interval), which is the point made in 
\cite{coussaert/henneaux93} to deem this transformation as non-gauge, 
and yet both metrics must be physically identified.  This result 
proves that we must fix the gauge even for the three-space 
diffeomorphisms, namely, the residual gauge group.  It is in this 
place that the HPD play the significant role we have seen in Section 
\ref{sec:4step} and Section \ref{sec:hamiltonian}.

Our analysis agrees with that of \cite{ashtekar/samuel91} with regard 
to the Lagrangian formulation, and we agree with their spacetime 
counting of the degrees of freedom (see the table in 
\cite{ashtekar/samuel91}; also see 
\cite{koike/tanimoto/hosoya94,tanimoto/koike/hosoya97a,%
tanimoto/koike/hosoya97b} for a counting of degrees of freedom in 
models with compactified spatial sections).  But we differ in other 
respects.  Let us make the differences clear; they concern the status 
of the HPD associated with ``outer'' automorphisms of the right 
invariant Lie algebra.

In \cite{ashtekar/samuel91}, ``outer'' HPD, that is, HPD yielding 
automorphisms of the Lie algebra that are not inner automorphisms, 
count as gauge degrees of freedom in the Lagrangian formulation, but 
are considered non-gauge symmetries in phase space.  In 
\cite{coussaert/henneaux93}, these ``outer'' HPD are always taken in 
both formulations as non-gauge symmetries.  Instead, from our 
systematic procedure of reducing the gauge group, we deduce that 
the ``outer'' HPD are gauge transformations in both the Lagrangian and 
the Hamiltonian formalisms, and they always count as gauge degrees of 
freedom.  

Also see our comments in Appendix \ref{app:freedom}.  For example, we 
explicitly consider the case of the Bianchi Type I model, where the 
number of degrees of freedom is 1 (if the surfaces of homogeneity have 
the topology of ${\rm I\kern-2pt R}^3$).  It may appear to some people 
that an odd number of degrees of freedom in a Hamiltonian formalism is 
somehow not correct, but that is the case here, and we discuss this 
matter in a bit more detail in Appendix \ref{app:freedom}.

Summing up, we have exhibited in a simple way the two different 
problems that appear when we implement the Killing conditions and the 
gauge fixing into the Lagrangian or into the Hamiltonian.  The first 
problem prevents the Class B Bianchi models from having Lagrangian or 
Hamiltonian formulations (except possibly in special cases), whereas 
the second is solved by introducing some requirements on the initial 
conditions.  We have also shown that there is no ambiguity or matter 
of interpretation in what must be understood as a gauge 
transformation, either in the Lagrangian or the Hamiltonian formalism.  
Finally our analysis proves that the number of degrees of freedom in 
both formalisms is always the same.


\section*{Acknowledgments}

J.M.P.\ would like to thank the Center for Relativity of The 
University of Texas at Austin for its hospitality.  J.M.P.\ 
acknowledges support by CICYT contracts AEN95-0590 and GRQ 93-1047 and 
wishes to thank the Comissionat per a Universitats i Recerca de la 
Generalitat de Catalunya for a grant.


\appendix
\section{General spatially homogeneous metric}
\label{app:homog}

In order to demonstrate some of our ideas explicitly, we here give 
formulas for the connection, curvature, and field equations for a 
general spatially homogeneous metric, including a general lapse 
function and shift vector.  A word on calculational procedure:  It is 
easiest to use an orthonormal basis adapted to the Killing structure:
\begin{equation}  
    {\bf g}
    = \eta_{\mu\nu}\bbox{\sigma}^{\mu}\bbox{\sigma}^{nu} 
    = - (\bbox{\sigma}^0)^2
    + \delta_{ij}\bbox{\sigma}^i\bbox{\sigma}^j\ ,
        \label{A.orthonorm}
\end{equation}
where the basis is defined by
\begin{equation}  
    \bbox{\sigma}^0 = N{\bf d}t\ ,\
    \bbox{\sigma}^i = b^i_a(\bbox{\omega}^a + N^a{\bf d}t)\ ,
            \label{A.basis}
\end{equation}
where $N,N^{a},g_{ab}$ are functions only of $t$, and we use the 
convention that bold face denotes a tensor or a form; the one-forms 
$\bbox{\omega}^{a}$ obey the following relation, indicating that they 
are invariant under the isometry group whose lie algebra is defined by 
the structure constants $C^{a}_{bc}$:
\begin{equation}  
    {\bf d}\bbox{\omega}^a 
        = {1\over2} C^a_{bc}\bbox{\omega}^b\wedge\bbox{\omega}^c\ .
            \label{A.omegas}
\end{equation}
Note that the three-metric is defined by
\begin{equation}  
    g_{ab} = \delta_{ij} b^i_a b^j_b\ ,
            \label{A.3metric}
\end{equation}
and we will consistently use $i,j,\dots$ for orthonormal indices, and 
$a,b,c,\dots$ for ordinary three-indices.  The matrix 
${\bf B}=(b^{i}_{a}(t))$ represents an arbitrary square root of the 
metric; precisely what form it takes will be irrelevant.  The inverse 
of the three-metric is defined by
\begin{equation}
    e^{ac}g_{cb}=\delta^{a}_{b}
\end{equation}
in order not to confuse it with the $ab$ components of the 
contravariant four-metric.  The inverse of {\bf B} is the matrix 
${\bf A}=(a^{a}_{i})$, so that
\begin{equation}  
    e^{ab} = a^a_i a^b_j \delta^{ij}\ ,\
    {\rm where}\ 
    a^a_j b^i_a = \delta^i_j\ \Longleftrightarrow
            \ a^a_i b^i_b = \delta^a_b\ .
                \label{A.inversemetric}
\end{equation}

The first Cartan equations with the torsion set equal to zero are
\begin{equation}
	{\bf d}\bbox{\sigma}^{\mu}
	  = - \bbox{\sigma}^{\sigma}\wedge\bbox{\sigma}^{\mu}_{\sigma}\ ,
	        \label{A.Cartan1}
\end{equation}
where $\bbox{\sigma}^{\mu}_{\nu}$ are the connection one-forms, used 
in forming the covariant derivative of a tensor.  The metric 
compatibility equations (covariant derivative of the metric equals 
zero), when the metric components are constants, as in an orthonormal 
basis, here read
\begin{equation}
    \eta_{\mu\sigma}\bbox{\sigma}^{\sigma}_{\nu}
    =-\eta_{\nu\sigma}\bbox{\sigma}^{\sigma}_{\mu}\ .
             \label{A.metriccompat}
\end{equation}
Equations (\ref{A.Cartan1}) and (\ref{A.metriccompat}) uniquely 
determine the connection one-forms.  The second Cartan equations are
\begin{equation}
	{1\over2}R^{\mu}_{\nu\sigma\tau}
	    \bbox{\sigma}^{\sigma}\wedge\bbox{\sigma}^{\tau}
   = {\bf d}\bbox{\sigma}^{\mu}_{\nu}
	+\bbox{\sigma}^{\mu}_{\sigma}\wedge\bbox{\sigma}^{\sigma}_{\nu}\ ;
	        \label{A.Cartan2}
\end{equation}
they determine the Riemann tensor components (see 
\cite{ryan/shepley75} for more details).  

It is convenient to define a matrix related to the logarithmic 
derivative of {\bf B} and its symmetric and antisymmetric parts as the 
matrices {\bf K}, {\bf L}, {\bf M}.  It is also convenient to define 
the orthonormal projection of the structure constants (which then 
become a time-dependent array {\bf D} plus another array {\bf E} by 
(here $\dot{}$ denotes $d/dt$):
\begin{mathletters}
	\begin{equation} 
	    K_{ij} = {1\over N}(\dot b^i_a a^a_j
		    - b^i_aN^ba^c_jC^a_{bc})\ ,
	\label{A.KLMdefa} 
	\end{equation}
	\begin{equation}
	    L_{ij} = {1\over2}(K_{ij}+K_{ji})\ ,\
	    M_{ij} = {1\over2}(K_{ij}-K_{ji})\ ,
	\label{A.KLMdefb}
	\end{equation}
	\label{A.KLMdef}
\end{mathletters}
\begin{equation}
    D^{i}_{jk} = b^i_aa^b_ja^c_kC^a_{bc}\ ,\
    E_{ijk} = D^i_{jk}-D^j_{ik}-D^k_{ij}\ .
        \label{A.DEdef}
\end{equation}
With these definitions, we have for the curls of the orthonormal 
basis forms
\begin{equation}
    {\bf d}\bbox{\sigma}^{0}=0\ ,\
    {\bf d}\bbox{\sigma}^{i}
        =K_{ij} \bbox{\sigma}^{0}\wedge\bbox{\sigma}^{j}
            + {1\over2}D^{i}_{jk}
                \bbox{\sigma}^{j}\wedge\bbox{\sigma}^{k}\ .
                   \label{A.curlsK}
\end{equation}
The connections forms are:
\begin{equation}
    \bbox{\sigma}^0_i=\bbox{\sigma}^i_0
      = L_{ij}\bbox{\sigma}^j\ ,\
    \bbox{\sigma}^i_j
      = - M_{ij}\bbox{\sigma}^0 + {1\over2} E_{ijk}\bbox{\sigma}^k\ .
          \label{A.connectK}
\end{equation}

The results for the independent components of the Riemann tensor are 
most conveniently displayed after raising an index (using 
$\delta^{ij}$):
\begin{mathletters}
\begin{eqnarray}  
    R^{0i}_{0j} 
    &=& {1\over N}\dot L_{ij} 
        +L_{ik}L_{jk} -L_{ik}M_{jk} -L_{jk}M_{ik}\ ,  
    \label{A.riemanna}\\
    R^{0i}_{k\ell}
    &=& L_{ij}D^j_{k\ell} 
        +{1\over2} (L_{jk}E_{ji\ell} - L_{j\ell}E_{jik})\ , 
    \label{A.riemannb}\\
    R^{ij}_{k\ell}
    &=& L_{ik}L_{j\ell}-L_{i\ell}L_{jk}
        \nonumber \\
    &-&{1\over4}E_{imk}E_{jm\ell}+{1\over4}E_{im\ell}E_{jmk}
        +{1\over2} E_{ijm}D^m_{k\ell}\ .
    \label{A.riemannc}
\end{eqnarray}
\label{A.riemann}
\end{mathletters}

The Ricci tensor components, the scalar curvature, and the Einstein
tensor components are defined by
\[ 
    R^{\mu}_{\nu}=R^{\mu\sigma}_{\nu\sigma}\ ,
    R= R^{\sigma}_{\sigma}\ ,
    G^{\mu}_{\nu}=R^{\mu}_{\nu}-{1\over2}R\delta^{\mu}_{\nu}\ .
\]
The evolution equations (in vacuum) are setting the space-space
components of the Ricci tensor to zero, where
\begin{eqnarray}  
    R^i_j &=& {1\over N}\dot L_{ij} -L_{ik}M_{jk} -L_{jk}M_{ik} 
            + L_{kk}L_{ij}    \nonumber \\
          &~& + {1\over2}(D^i_{jk} + D^j_{ik})D^\ell_{k\ell} 
               \nonumber \\
          &~& - {1\over2}D^\ell_{ik}(D^k_{j\ell} + D^\ell_{jk}) 
              + {1\over4}D^i_{k\ell}D^j_{k\ell}\ .
    \label{A.Rij}
\end{eqnarray}
Next we display the time-time and time-space components of the 
Einstein tensor; these set to zero are the constraint equations on 
initial value data:
\begin{mathletters}
	\begin{eqnarray}
	    G^{0}_{0} &=& {1\over2}L_{st}L_{st} 
	                    - {1\over2}(L_{ss})^2 
	                \nonumber \\
	        &~&  + {1\over2}D^{t}_{st}D^{u}_{su} 
	             +{1\over4}D^{t}_{su}D^{u}_{st}
	             +{1\over8}D^{s}_{tu}D^{s}_{tu} \ ,
	                \label{A.G00x} \\
	    G^0_i   &=& -L_{ij}D^k_{jk} - L_{jk}D^j_{ki}\ .
	                \label{A.G0i}
	\end{eqnarray}
	\label{A.G000}
\end{mathletters}

Finally, we will need the scalar curvature in order to display the 
reduced Lagrangian.  In the action integral the volume element is 
really
\begin{equation}  
    \bbox{\sigma}^o\wedge\bbox{\sigma}^1\wedge
        \bbox{\sigma}^2\wedge\bbox{\sigma}^3
    = N\sqrt{g}\,
        {\bf d}t\wedge\bbox{\omega}^1\wedge
        \bbox{\omega}^2\wedge\bbox{\omega}^3\ ,
\end{equation}
where $g=\det(g_{ab})$, and the spatial integral can be set to the 
constant $V$:
\begin{equation}  \int\int\int\bbox{\omega}^1\wedge
        \bbox{\omega}^2\wedge\bbox{\omega}^3 = V\ .
\end{equation}
The action integral is therefore
\[  {\cal I}= \int {\cal L} V\,dt\ ,
\]
where
\begin{eqnarray}
	{\cal L} & = & R\,N\,\sqrt{g}
	\nonumber  \\
	 & = & (2\sqrt{g}\,L_{ss})\,\dot{}
	     -2\sqrt{g}N^{a}b^{s}_{a}D^{u}_{su}L_{ss}
	\nonumber  \\
	 &  & + N\sqrt{g}\bigg(L_{st}L_{st} - (L_{ss})^{2}
	\nonumber \\
	 &  & \quad  -D^{t}_{st}D^{u}_{su}
       -{1\over2}D^{t}_{su}D^{u}_{st}
       -{1\over4}D^{s}_{tu}D^{s}_{tu}\bigg) \ ,
	\label{A.lagrg}
\end{eqnarray}
and note that we have separated out a total time derivative.  Here it 
is best to be explicit, and we display the Lagrangian in terms of the 
configuration space variables $N,N^{a},g_{ab}$:
\begin{eqnarray}  
    {\cal L} &=&
    {\sqrt{g}\over N}\bigg[
        -{1\over4}(e^{ab}\dot g_{ab})^2
        +{1\over4}\dot g_{ab} \dot g_{cd} e^{ac} e^{bd}    
                  \nonumber \\
    &~&\qquad -\dot g_{ab} e^{bc} N^d C^a_{dc}
        + (N^aC^b_{ab})^2
                  \nonumber \\
    &~&\qquad +{1\over2} g_{ab}e^{cd}N^eN^fC^a_{ec}C^b_{fd}
       +{1\over2} N^eN^fC^a_{eb}C^b_{fa}  \bigg]  
                  \nonumber \\
    &~&-N\sqrt{g}\bigg[
        + e^{bc}C^a_{ba}C^d_{cd} 
        + {1\over2} e^{cd}C^a_{cb}C^b_{da}
                  \nonumber \\
    &~&\qquad\qquad\qquad
         +{1\over4} g_{ad}e^{be}e^{cf}C^a_{bc}C^d_{ef} \bigg] \ .
\end{eqnarray}

The vacuum field equations are most conveniently written as four 
constraint equations and six evolution equations:
\begin{equation}
    G^{0}_{0}=0\ , G^{0}_{i}=0\ ,
    R^{i}_{j}=0\ .
\end{equation}
These equations can be derived from the Einstein-Hilbert Lagrangian 
density before imposing any symmetry requirements and then imposing 
the symmetry requirements after the equations have been derived.  We 
will compare these equations with those derived from the reduced 
Lagrangian $\cal L$; in other words, by imposing the symmetry 
requirements first and then deriving equations.

The equation $\delta{\cal L}/\delta N=0$ is readily seen to be
exactly $G^0_0=0$ (up to a factor of $2\sqrt{g}$).

The equations $\delta{\cal L}/\delta N^a=0$ (after canceling $N$
and $\sqrt{g}$) are
\begin{eqnarray} 
    0 &=& +\dot g_{db}g^{bc}C^d_{ca}
        +2C^c_{ac}N^bC^d_{bd}
            \nonumber \\
    &~&~~+g_{ef}g^{cd}N^bC^f_{bdl}C^e_{ac}
    +N^bC^d_{bc}C^c_{ad}\ .
\end{eqnarray}
This is not proportional to the $G^0_i$ equation in a Class B model; 
the difference (up to nonzero factors) is
\begin{equation}  
    \left[\dot g_{ab}g^{bf} +2C^c_{ac}N^f 
        -g_{ae}g^{cf}N^bC^e_{bc}\right]
        C^d_{fd} \ .
                   \label{A.extra}
\end{equation}

In other words, if the lapse is kept in, then all of the constraint 
equations can be derived for the Class A models.  In Class B, if these 
terms happen to be zero anyway, well and good, but they won't 
automatically vanish.  Note several things about the terms in equation 
(\ref{A.extra}): The first one is independent of $N^a$; therefore, one 
might hope that adding an appropriate term to the Lagrangian might 
remove it (a term, that is, linear in $N^a$).  The second term has 
$N^a$ multiplied by a matrix which is symmetric; supposedly it too 
could be removed with an appropriate term in the Lagrangian.  The last 
term, however, has $N^a$ times a matrix which is not symmetric; it 
would seem that there is no hope of generating this term by adding 
something to the Lagrangian.


\section{Spatially homogeneous Maxwell theory}
\label{app:maxwell}

A relatively simple system which illustrates many of our ideas is that 
of a spatially homogeneous electromagnetic potential which obeys the 
Maxwell equations.  (Here we start with the four-vector potential 
rather than with a spatially homogeneous field tensor, which of course 
could be produced using a non-homogeneous potential vector.)  The 
background metric is taken to be a simple spatially homogeneous one, 
so we do not require that it obey any particular field equations.  We 
take the metric to have components $\eta_{\mu\nu}$ in an invariant 
basis:
\begin{equation}
    {\bf g}= -{\bf d}t^{2} 
        +\delta_{ij}\bbox{\omega}^{i}\bbox{\omega}^{j}\ ,
            \label{B.metric}
\end{equation}
where
\begin{equation}
    {\bf d}\bbox{\omega}^{i}
        ={1\over2}C^{i}_{jk}
            \bbox{\omega}^{j}\wedge\bbox{\omega}^{k}\ .
                \label{B.commrlns}
\end{equation}
Note that there is a freedom to transform the spatial basis 
$\{\bbox\omega^{i}\}$ by an orthogonal transformation but not 
necessarily by a general linear transformation.  

The vector potential actually is an equivalence class of one-forms 
related by gauge transformations.  Here we demand that the class 
contain at least one member which is spatially homogeneous.  If that 
one is expressed in the invariant basis, its components are functions 
only of $t$:
\begin{equation}
	{\bf A} = A_{0}(t){\bf d}t +A_{i}(t)\bbox{\omega}^{i}\ ,
	\label{B.vecpot}
\end{equation}
and it produces the following field 2-form:
\begin{equation}
	{\bf F}={\bf dA}= \dot A_{i}{\bf d}t\wedge\bbox{\omega}^{i}
	    +{1\over2}A_{i}C^{i}_{jk}
            \bbox{\omega}^{j}\wedge\bbox{\omega}^{k}\ .
	\label{B.field}
\end{equation}
Note that $A_{0}$ has disappeared; it plays no part either in the 
Maxwell equations or in the Lagrangian.  It can be made to vanish by a 
gauge transformation of the kind
${\bf A}\longrightarrow{\bf A}'={\bf A}+{\bf d}\lambda(t)$.

We choose to concentrate on the metric as expressed in an invariant 
basis; this is a choice of metric gauge.  Maxwell theory cannot be 
expressed without at least some reference to a background spacetime 
metric, and so this metric gauge affects the electromagnetic 
potential.  We also choose to concentrate on the vector potential 
within its equivalence class which explicitly has vanishing Lie 
derivative with respect to the generators of the invariance group.  
This, too, is a choice of gauge, though as we discuss below, in some 
cases there is still some residual gauge freedom.  

The easiest way to calculate the Maxwell equations (in vacuum) is 
first to write down the dual field 2-form:
\begin{equation}
	{}^{*}{\bf F}={1\over2}\epsilon_{ijk}A_{s}C^{s}_{jk}
	        {\bf d}t\wedge\bbox{\omega}^{i}
	    -{1\over2}\epsilon_{ijs}\dot A_{s}
	        \bbox{\omega}^{i}\wedge\bbox{\omega}^{j}\ ,
	\label{B.dualfield}
\end{equation}
where $\epsilon_{ijk}$ is the Levi-Civita symbol, equal to $\pm1$ if 
($ijk$) is an even/odd permutation of (123) and to zero otherwise.  
The Maxwell equations are:
\begin{eqnarray}
	0&=&{\bf d}^{*}{\bf F}  \nonumber \\
	&=& -{1\over2}\left(\epsilon_{stu}\ddot A_{s}
	        +{1\over2}\epsilon_{ijk}A_{s}C^{s}_{jk}C^{i}_{tu}\right)
	    {\bf dt}\wedge\bbox{\omega}^{t}\wedge\bbox{\omega}^{u}
	        \nonumber \\
	&~&~~-{1\over2}\epsilon_{isk}\dot A_{k}C^{i}_{tu}
	  \bbox{\omega}^s\wedge\bbox{\omega}^t\wedge\bbox{\omega}^u\ ,
\end{eqnarray}
and they are equivalently written as:
\begin{mathletters}\label{B.maxwell}
\begin{equation}
	0 = \dot A_{j}C^{s}_{js}\ ,
	\label{B.maxwell.0}
\end{equation}
\begin{equation}
	0 = \ddot A_{i}+ {1\over2}A_{s}C^{s}_{jk}C^{i}_{jk}
	    + A_{s}C^{s}_{it}C^{u}_{tu}\ .
	\label{B.maxwell.i}
\end{equation}
\end{mathletters}%
One can double-check these equations by forming the connection 
one-forms and computing $F^{\mu\sigma}\!_{;\sigma}=0$.

The action integral for a general Maxwell vector potential is
(${}^{4}g$ is the determinant of the spacetime metric in
a coordinate system):
\begin{equation}
	{\cal I} 
	= \int {1\over4}F^{\mu\nu}F_{\mu\nu}\sqrt{|{}^{4}g|}\,d^{4}x \ .
	\label{B.action}
\end{equation}
In this case, $\sqrt{|{}^{4}g|}$ is independent of $t$, and so the 
reduced Lagrangian ${\cal L}_{R}$ is:
\begin{equation}
	{\cal L}_{R}={1\over4}F^{\sigma\tau}F_{\sigma\tau}
	= -{1\over2}\dot A_{s} \dot A_{s}
	    +{1\over4} A_{s} A_{t}
	        C^{s}_{jk}C^{t}_{jk}\ .
	\label{B.lagrangian}
\end{equation}
The Euler-Lagrange equations 
$0=\delta{\cal L}_{R}/\delta A_{i}$ are:
\begin{equation}
	0 = \ddot A_{i}+{1\over2}A_{s}C^{i}_{jk}C^{s}_{jk}\ .
	\label{B.ELeqns}
\end{equation}

It is clear that in Class B models equations (\ref{B.maxwell}) and 
(\ref{B.ELeqns}) do not agree.  There are two differences: Equation 
(\ref{B.maxwell.0}) is a constraint equation, and it simply cannot 
come from a variational principle which is homogeneous quadratic in 
the velocities.  Equation (\ref{B.maxwell.i}) has the term 
$A_{s}C^{s}_{it}C^{u}_{tu}$ which does not appear in equation 
(\ref{B.ELeqns}).  Notice that equation (\ref{B.maxwell.0}) is 
automatically satisfied in the Class A case ($C^{s}_{js}=0$), and also 
equations (\ref{B.maxwell.i}) and (\ref{B.ELeqns}) then do agree.

One thing which is true, however, is that the constraint 
equation (\ref{B.maxwell.0}) is compatible both with equation 
(\ref{B.maxwell.i}) and equation (\ref{B.ELeqns}).  To see this fact, 
take the time derivative of equation (\ref{B.maxwell.0}):
\[
	0 = \ddot A_{j}C^{s}_{js}\ .
\]
When equation (\ref{B.maxwell.i}) is multiplied by $C^{s}_{is}$, the 
result is
\[
	0 = C^{s}_{is}\ddot A_{i} + {1\over2}A_{t}C^{t}_{jk}C^{s}_{is}
	  + A_{j}C^{j}_{it}C^{u}_{tu}C^{s}_{is}\ .
\]
The middle term vanishes because of the Jacobi identity, which in a 
three-dimensional Lie algebra is equivalent to $0 = 
C^{s}_{ij}C^{t}_{st}$.  The last term, which would be absent anyway if 
equation (\ref{B.ELeqns}) had been used, vanishes as a consequence of 
the antisymmetry of $C^{j}_{it}$ in its lower indices.

Thus, even if the constraint equation (\ref{B.maxwell.0}) were put in 
by hand, the problem would remain whether any variational principle 
could reproduce the last term in equation (\ref{B.maxwell.i}).

We now turn to gauge transformations.  A gauge transformation here is 
the addition to the vector potential of a homogeneous one-form 
 whose curl vanishes (and therefore which can at least 
locally be expressed as the curl of a function).  Let this one-form be
\begin{equation}
	\bbox\kappa = \kappa_{0}{\bf d}t + \kappa_{i}\bbox\omega^{i}\ ,
	\label{B.defkappa}
\end{equation}
where the components $\kappa_{\mu}$ are functions only of $t$.  We 
require that the curl of $\bbox{\kappa}$ be zero:
\begin{equation}
	0 = {\bf d}\bbox{\kappa}
	  = \dot\kappa_{i}{\bf d}t\wedge\bbox\omega^{i}
	    + {1\over2}\kappa_{i}C^{i}_{jk}
	        \bbox\omega^{j}\wedge\bbox\omega^{k}\ .
	\label{B.kappacurl}
\end{equation}
Therefore $\kappa_{0}$ is arbitrary, and $\kappa_{i}$ must be a set of 
constants subject to the condition
\begin{equation}
	\kappa_{i}C^{i}_{jk} = 0\ .
	\label{B.kappacondition}
\end{equation}
Only Bianchi Types VIII and IX require that $\kappa_{i}$ be zero.  In 
fact, in a Class B model, $\kappa_{i}$ may be taken to be 
proportional to $C^{j}_{ij}$. 

We illustrate with the example of the general Class B model, in which 
only $C^{s}_{1s}\ne0$ (this prescription can always be satisfied in 
any Class B model by using an orthogonal transformation of the 
invariant basis).  The Jacobi identity then requires that 
$C^{1}_{ij}=0$.  The reduced Lagrangian in this case is
\begin{equation}
	{\cal L}_{R} 
	= -{1\over2}(\dot A_{1}^{2} +\dot A_{2}^{2} +\dot A_{3}^{2})
	        + {1\over4}A_{A}A_{B}C^{A}_{jk}C^{B}_{jk} \ ,
	\label{B.redlagr}
\end{equation}
where the indices $A,B$ range only over (2,3).  The Euler-Lagrange 
equations are
\[ 
    0 = \ddot A_{1}\ ,\
    0 = \ddot A_{A} + {1\over2}A_{B}M_{AB} \ ,
\]
where
\[ 
    M_{AB}\equiv C^{A}_{jk}C^{B}_{jk} \ .
\]
The Maxwell Equations (\ref{B.maxwell}) read
\[ 
    0 = \dot A_{1}\ ,
\]
\[ 
    0 = \ddot A_{1}\ ,\
    0 = \ddot A_{A} + {1\over2}A_{B}M_{AB} 
        + A_{B}N_{AB} \ , 
\]
where 
\[ 
    N_{AB} \equiv C^{B}_{A1}C^{s}_{1s} \ .
\]  

The reduced Maxwell equations thus differ in two ways from the 
Euler-Lagrange equations: First, the Maxwell equations have the 
constraint equation $\dot A_{1}=0$.  The residual gauge freedom 
allowed by equation (\ref{B.kappacondition}) says that $A_{1}$ can be 
made zero by a choice of gauge.  In contrast, the Euler-Lagrange 
equations only require that $A_{1}$ be at most a linear function of 
$t$, and the gauge freedom only will allow the value of $A_{1}$ to be 
set to zero at a particular time.  It is not, however, difficult to 
add the constraint $\dot A_{1}=0$ to the Euler-Lagrange equations in 
an \emph{ad hoc} manner.

Second, the Maxwell equations for $A_{A}$ have the additional term 
involving $N_{AB}$.  If it happens that $N_{AB}$ is symmetric, then a 
Lagrangian can be found to reproduce this term: The Lagrangian would 
add the term ${1\over2}A_{A}A_{B}N_{AB}$ to the reduced Lagrangian 
equation (\ref{B.redlagr}).  However, if $N_{AB}$ is not symmetric, 
then such a Lagrangian would not in general be possible --- certainly 
no Lagrangian could be found to reproduce the Maxwell equations 
exactly, though under some circumstances it may be possible to find a 
Lagrangian which would produce equations equivalent to the Maxwell 
equations.

For example, the standard structure constants of a Bianchi Type III 
model \cite{taub51} are $C^{2}_{12}=-C^{2}_{21}=1$, the other 
structure constants vanishing.  In this case $N_{AB}={\rm diag}(1,0)$, 
and the Maxwell equations can indeed be derived from a variational 
principle (with the constraint $0=\dot A_{1}$ being put in by hand): 
The Maxwell equations are:
\[ 
    0 = \dot A_{1}=\ddot A_{1}=\ddot A_{2}=\ddot A_{3}\ ,
\]
which clearly can come from a constrained variational principle.  Note 
that in this Type III case the reduced Lagrangian yields as 
Euler-Lagrange equations:
\[ 
    0=\ddot A_{1}=\ddot A_{2}+A_{2}=\ddot A_{3}\ ,
\]
which are not at all the same as the Maxwell equations.

A second example is that of a group which has structure constants 
$C^{2}_{12}=-C^{2}_{21}=C^{2}_{13}=-C^{2}_{31}=1$, the rest being 
zero.  In fact, this group is also Type III, in a basis which is a 
linear transformation of the basis in the preceding example.  Since 
this transformation is not an orthogonal one, the Maxwell equations in 
this case differ significantly from the preceding.  The reduced 
Lagrangian is
\[
     {\cal L}_{R} 
     = -{1\over2}(\dot A_{1}^{2} +\dot A_{2}^{2} +\dot A_{2}^{2})
         + A_{2}^{2}\ .
\]
The Euler-Lagrange equations are:
\[
    0 = \ddot A_{1} = \ddot A_{2} + 2A_{2} = \ddot A_{3} \ .
\]
The Maxwell equations are:
\[
    0 = \dot A_{1}=\ddot A_{1}=\ddot A_{2}=\ddot A_{3}-A_{2}\ .
\]
The difference between the Maxwell equations and the Euler-Lagrange 
equations in this example are profound:  First, the Maxwell equations 
include the constraint $\dot A_{1}=0$, as in the previous example.  
Second, the Maxwell equations cannot be derived from a variational 
principle, unlike the previous example.  


\section{Relationship between setting the gauge and variational
principles}
\label{app:varprin}

Here we summarize and expand the main result of \cite{pons96}.  For 
simplicity we will use the language of mechanics (finite number of 
degrees of freedom), although everything can be translated to field 
theory.  Consider a singular Lagrangian $L$ that leads to some primary 
constraints, $\phi_\rho = 0$ in phase space, presumed to be effective 
(each has non-vanishing gradient on the constraint surface).  For the 
sake of simplicity we will consider the case when all the constraints, 
primary, secondary, etc., are first class (this is what happens in our 
generally covariant theories).  The equations of motion obtained from 
$L$ are:
\[  {\delta L\over\delta q^{i}} = \alpha_i - W_{ij}\ddot q^j = 0\ , 
\]
where
\[  W_{ij}\ddot q^j 
    = {\partial^{2}L\over\partial\dot q^{i}\dot q^{j}}\ ,\ 
    {\rm and}\
    \alpha_i
    ={\partial L\over\partial q^{i}}
        - {\partial^{2}L
            \over\partial\dot q^{i}\partial q^{j}}\dot q^{j}
        - {\partial^{2}L\over\partial\dot q^{i}\partial t} \ .
\]

The Lagrangian dynamics can equivalently be described by a vector 
field that exists on, and is tangent to, the constraint surface in 
configuration-velocity space:
\begin{equation}
    {\bf X} \equiv {\partial\over\partial t}
        + \dot q^i {\partial\over\partial q^i}
        + a^i(q,\dot q){\partial\over\partial\dot q^i} 
        +\eta^\rho\bbox{\Gamma}_\rho 
    \equiv {\bf X}_0 + \eta^\rho\bbox{\Gamma}_\rho\ .
                                               \label{C.lag-ev}
\end{equation}
The $a^i$ are determined from the equations of motion and the 
stabilization algorithm; $\eta^\rho$ are arbitrary functions of time 
(or spacetime in field theory) and any other variable; and  
$\bbox{\Gamma}_\rho$ is
\[ 
    \bbox{\Gamma}_\rho = \gamma^i_\rho 
	{\partial\over\partial\dot q^i}\ , 
\]
where $\gamma^i_\rho$ are the null vectors of $W_{ij}$. These null 
vectors can be given as
\begin{equation} 
    \gamma^i_\rho = {\partial\phi_\rho\over\partial p_i}(q,\hat p)\ , 
                                                \label{C.gamma}
\end{equation}
where $\hat p_i(q,\dot q)=\partial L/\partial\dot q^i$.  Let us point 
out that $\eta^\rho\bbox{\Gamma}_\rho$ in equation (\ref{C.lag-ev}) is 
the piece that corresponds in our case to equation (\ref{freed}).

Notice that $\alpha_i \gamma^i_{\rho} = 0$ is a consequence of the 
equations of motion $\delta L/\delta q^{i} = 0$.  They are called the 
primary Lagrangian constraints.

Consider now a partial gauge-fixing of the dynamics given by a set of 
new constraints $\chi_{\rho'} = 0$, with $|\rho'| < |\rho|$, defined 
in configuration space (holonomic constraints).  Let us split the set 
of indices $\rho$ into two sets, $\rho'$ and $\rho''$, in such a way 
that $\bbox{\Gamma}_{\rho''} {\dot \chi}_{\sigma'} = 0$ and 
$|\bbox{\Gamma}_{\rho'} {\dot \chi}_{\sigma'}| \neq 0$.  Then, the 
requirement ${\bf X} ({\dot \chi}_{\rho'}) = 0$ determines the 
functions $\eta^{\rho'}$ and leaves completely undetermined 
$\eta^{\rho''}$.

In such a situation, the following result is proved in \cite{pons96}:
If we plug the holonomic gauge-fixing $\chi_{\rho'} = 0$ into the 
original Lagrangian $L$ to get the reduced Lagrangian $L_{\rm GF}$, 
then the following equivalence holds (where $[L]$ is the 
Euler-Lagrange variation):
\begin{equation}
    [L] = 0 \ , \ \chi_{\rho'} = 0
          \Longleftrightarrow
    [L_{\rm GF}] = 0 \ , \ \alpha_i \gamma^i_{\rho'} = 0 \ , 
                                     \label{C.big-equiv} 
\end{equation}
where $\alpha_i \gamma^i_{\rho'}$ are a subset of the primary 
Lagrangian constraints for $L$ with the understanding that the gauge 
fixing has been plugged into them.

So we see that to describe the same motions in the reduced space it 
is not enough in general to impose the new Lagrangian equations 
of motion, $[L_{\rm GF}] = 0$, but some additional constraints must 
be required, too.

Now, we will prove, using the machinery of \cite{pons96}, that 
in order for the solutions of $[L_{\rm GF}] = 0$ to satisfy the 
constraints $\alpha_i \gamma^i_{\rho'} = 0$, we only need to impose 
them in the initial conditions.

Since the gauge-fixing constraints $\chi_{\rho'} = 0$ are holonomic, 
they just reduce the configuration space. We have adopted the 
notation $q^i$ for the local coordinates in the original 
configuration space. We will adopt the notation $Q^a$ for the local 
coordinates in the reduced configuration space. 

It is proved in \cite{pons96} that the Lagrangian evolution operator 
in the reduced velocity space takes the form
\begin{eqnarray}
    {\bf X_{\rm R}}&\equiv& {\partial\over\partial t} 
               + \dot Q^a {\partial\over\partial Q^a}
               + a^a(Q,\dot Q){\partial\over\partial\dot Q^a} 
               +\tilde{\eta}^{\rho''}\tilde{\bbox{\Gamma}}_{\rho''}
          \nonumber \\
     &\equiv& {\bf X_{\rm R}}_0 + 
               \tilde\eta^{\rho''}\tilde{\bbox{\Gamma}}_{\rho''}\ ; 
                                     \label{C.lagred-ev} 
\end{eqnarray}
where $\tilde\eta^{\rho''}$ are arbitrary functions; 
$\tilde{\bbox{\Gamma}}_{\rho''}$ is
\[ 
    \tilde{\bbox{\Gamma}}_{\rho''}
    = \tilde{\gamma}^a_{\rho''} {\partial\over\partial\dot Q^a}\ ; 
\]
and similarly to equation (\ref{C.gamma}), $\tilde{\gamma}^a_{\rho''}$ 
is defined by
\begin{equation} 
    \tilde{\gamma}^a_{\rho''}
    = {\partial\tilde{\phi}_{\rho''} \over\partial P_a}(Q,\hat P)\ , 
                                          \label{C.gamma-red}
\end{equation}
where $\tilde{\phi}_{\rho''}$ are the primary Hamiltonian constraints 
corresponding to the reduced theory. It turns out \cite{pons96} that 
these constraints are related to the original constraints 
$\phi_{\rho''}$ by
\begin{equation} 
    \phi_{\rho''}(q(Q),p)
    =\tilde{\phi}_{\rho''}(Q,p {\partial q\over\partial Q}) \ .
                                            \label{C.tilde}
\end{equation}

From all these results, the following equalities hold: 
\begin{eqnarray} 
    \tilde{\bbox{\Gamma}}_{\rho''} 
            \bigr((\alpha_i \gamma^i_{\rho'})|_{q(Q)}\bigr) 
     &=& {\partial \tilde{\phi}^a_{\rho''}\over\partial P_a} 
        {\partial\bigr((\alpha_i \gamma^i_{\rho'})|_{q(Q)}\bigr) 
                                 \over\partial \dot Q^a}
                                 \nonumber \\
    &=& ({\bbox{\Gamma}}_{\rho''} 
            \bigr(\alpha_i \gamma^i_{\rho'})\bigr)|_{q(Q)} 
                                 \nonumber \\
    &=& {{\cal F}\!L}^* \{\phi_{\rho'},\phi_{\rho''}\}|_{q(Q)}=0\ ,
                                 \nonumber \\
\end{eqnarray}
Here ${{\cal F}\!L}^*$ is the pullback of the Legendre map 
${{\cal F}\!L}$ from velocity space to phase space.  We have used in 
the last equality the fact that the pullback of a primary Hamiltonian 
constraint is identically zero.

With this new result, $\tilde{\bbox{\Gamma}}_{\rho''} \bigr((\alpha_i 
\gamma^i_{\rho'})|_{q(Q)}\bigr)=0$, we can see that the arbitrary part 
in the reduced Lagrangian evolution operator equation 
(\ref{C.lagred-ev}) has no effect on 
$(\alpha_i \gamma^i_{\rho'})|_{q(Q)}$.  This means 
that in the constraint surface for the reduced theory (this surface is 
determined by the equations of motion for $L_{\rm GF}$), the operator 
${\bf X_{\rm R}}_0$ must also be tangent to the surface 
$\bigr((\alpha_i \gamma^i_{\rho'})|_{q(Q)}\bigr) = 0$.  
(This is the only way to ensure the 
equivalence in equation (\ref{C.big-equiv}), for if the operator 
${\bf X_{\rm R}}_0$ is not tangent to the surface 
$\bigr((\alpha_i \gamma^i_{\rho'}\bigr)|_{q(Q)}) = 0$, 
then there will be no solutions of 
the equations of motion for the original Lagrangian in the gauge 
$\chi_{\rho'} = 0$, and we know that these solutions exist.)  
Therefore, if we consider a solution of $[L_{\rm GF}] = 0$ with 
initial conditions satisfying $(\alpha_i \gamma^i_{\rho'})|_{q(Q)}=0$, 
then the whole solution satisfies $(\alpha_i 
\gamma^i_{\rho'})|_{q(Q)}=0$.


\section{Degrees of Freedom}
\label{app:freedom}

In this Appendix we illustrate with a very simple case the counting of 
degrees of freedom for our models.  We also discuss the fact that 
spacetime rigid symmetries may be understood as residual gauge 
transformations.

We start by counting the true degrees of freedom for the Bianchi Type 
I case.  The counting can be done in configuration-velocity space or 
in phase space --- the counting is the same \cite{pons/shepley95}.  We 
start with the variables $N$ and ${\dot N}$, $N^{a}$ and $\dot N^{a}$, 
$g_{ab}$ and ${\dot g}_{ab}$:  20 apparent degrees of freedom.  The 
shift vector variables may be eliminated by our general 
considerations, so that 14 apparent degrees of freedom are left.

There is one constraint in the Lagrangian formalism.  To fix the time 
reparameterization invariance we define the time parameter as a 
function of our dynamical variables, excluding $N$.  Its stability, 
that is, the fact that the time derivative of this definition must 
vanish, will give a new constraint, containing the variable $N$, with 
no time dependence.  The stability of this last constraint gives a new 
constraint that can be used to isolate ${\dot N}$.  The requirement of 
stability now determines the arbitrary function in the Lagrangian 
evolution operator, and no more constraints appear.  We are left with 
1+3 constraints (one true constraint plus three gauge fixing 
constraints), lowering the number of degrees of freedom to 10.  The 
same counting can be done in phase space: In this case, there are 2 
true constraints, and we must introduce 2 gauge fixing constraints.

In Type I the momentum constraints are identically zero, so they do 
not reduce the degrees of freedom.  We are only left with HPD 
symmetries.  Every constant matrix $B^a_b$ of equation (\ref{7.16}) 
defines an automorphism of the Lie algebra.

Now topology enters the picture.  If the topology of $\Sigma$ is 
``open'' (for example, ${\rm I\kern-2pt R}^3$), simply connected, and 
with global Killing vectors, then to each matrix $B^a_b$ there 
corresponds an HPD through equation (\ref{hpd-y}).  Therefore, there 
are nine HPD gauge degrees of freedom.  The generators of HPD, 
equation (\ref{genhpd}), are not constraints, because in this case 
equations (\ref{gen3}) and (\ref{gen3ok}) differ by boundary 
terms.  So the final number of degrees of freedom for a Bianchi 
Type I model with surfaces of homogeneity with the topology of 
${\rm I\kern-2pt R}^3$ is $10-9=1$.

It may appear strange to some people that a Hamiltonian formalism can 
turn out to have an odd number of degrees of freedom, as in the above 
example.  In fact, we have already said that HPD need only be 
implemented in the initial conditions, so in this sense we have a 
Hamiltonian formulation plus an equivalence relation coming from the 
outset: It is not generated by the reduced formalism but is a remnant 
of the generally covariant theory we started with.  To find the 
number of degrees of freedom number is a matter of counting 
constraints, which trajectories count as the ``same'' as others, 
and whether a parameterized trajectory or its orbit (its 
one-dimensional set of points in phase space) should be 
considered as physically significant.  We have been guided by the 
principle that the Lagrangian and Hamiltonian formalisms should 
be equivalent (see \cite{pons/shepley95}) in coming to the 
conclusion that in fact they are.

If the topology of $\Sigma$ is that of a three-torus, then there are 
no infinitesimal HPD available (there are finite HPD but they do not 
reduce local degrees of freedom).  We end up with a final number of 10 
degrees of freedom.  We will not consider other topologies.

Let us turn to the second consideration in this Appendix, the 
possibility of interpreting rigid spacetime symmetries as residual 
gauge transformations.  Consider a free particle in 
${\rm I\kern-2pt R}^4$ in a gravitational background.  The action is
\begin{equation}
	S = \int \,(g_{\mu\nu} {\dot x}^\mu {\dot x}^\nu)^{1/2} d \tau\ ,
\end{equation}

What are the degrees of freedom?  If we start with a generally 
covariant theory, we know that diffeomorphisms correspond to gauge 
degrees of freedom.  Let us consider the metric background as 
non-dynamical, take the passive view for the action of 
diffeomorphisms, and consider the simplified case where there exists a 
system of spacetime coordinates such that the metric is just 
Minkowski.  Now we can make the following gauge fixing: We decide to 
stick with Minkowskian coordinates and only allow further 
diffeomorphisms if they keep this condition.  We end up with 
Poincar\'e transformations as residual gauge transformations.

To count the degrees of freedom, we start with positions $x^\mu$ and 
velocities ${\dot x}^\mu$, which amount to 8 degrees of freedom.  To 
fix the $\tau$ reparameterization invariance we must spend two 
constraints (for instance $x^0 - \tau = 0$ and its stabilization 
${\dot x}^0 = 1$).  We are left with 6 degrees of freedom.  The 
residual gauge freedom consists of Lorentz transformations, but our 
gauge fixing forbids boosts and time translations, so we are left with 
three-translations and rotations.  Rotations only affect 2 degrees of 
freedom, because the norm of the velocity three vector is unchanged, 
and therefore we eliminate 5 of the 6 remaining degrees of freedom.  
We end up with a single degree of freedom that in our gauge fixing 
corresponds to the kinetic energy.

Here we see a matter of interpretation as to what is and what is not a 
true degree of freedom.  From our point of view, if the Lorentz 
invariance may be considered as the residual invariance found after a 
process of reducing the gauge group of general covariance, then 
the Lorentz degrees of freedom must be considered as gauge.  But as 
residual gauge symmetries, they are not associated to constraints but 
to ordinary constants of motion.

Summing up, the analysis of degrees of freedom, either in the case of 
Bianchi cosmologies or the simple case of a Minkowskian free particle, 
depends upon the point of view adopted.  If one sticks to the 
formalism by itself, that is to say, to what the given Lagrangian 
yields as constraints, gauge transformations, and so on, one does not 
get the same number of degrees of freedom as if one considers that our 
theory comes from the reduction of an originally generally covariant 
theory.  In this second point of view (which is the one we support)
there are some symmetry 
transformations that are residual gauge transformations, that is, 
remnants of the reduction procedure of the Lagrangian and the gauge 
group, which require gauge fixing.



\end{multicols}


\begin{references}

\bibitem{taub51} 
A. H. Taub, 
    Ann. Math. {\bf 53}, 472 (1951) \vskip8pt  

\bibitem{ellis/maccalum69} 
G. F. R. Ellis and M. A. H. MacCallum, 
    Commun. Math. Phys. {\bf 12}, 108 (1969) \vskip8pt   

\bibitem{hawking69} 
S. W. Hawking,
    Mon. Not. R. Astron. Soc. {\bf 142}, 129 (1969) \vskip8pt   

\bibitem{maccallum71} 
M. A. H. MacCallum, 
    Commun. Math. Phys. {\bf 20}, 57 (1971) \vskip8pt   

\bibitem{maccallum/taub72} 
M. A. H. MacCallum and A. H. Taub, 
    Commun. Math. Phys. {\bf 25}, 173 (1972)  \vskip8pt  

\bibitem{ryan74} 
M. P. Ryan, 
    J. Math. Phys. (N.Y.) {\bf 15}, 812 (1974)  \vskip8pt  

\bibitem{sneddon76} 
G. E. Sneddon, 
    J. Phys. A: Math. Gen. {\bf 9}, 229 (1976)  \vskip8pt  

\bibitem{maccallum79} 
M. A. H. MacCallum, in S. W. Hawking and W. Israel, Editors, 
    {\it General Relativity, an Einstein centenary survey}, Chap. 11 
    (Cambridge University Press, Cambridge, 1979)  \vskip8pt  

\bibitem{ashtekar/samuel91} 
A. Ashtekar and J. Samuel, 
    Class. Quantum Grav. {\bf 8}, 2191 (1991)  \vskip8pt  

\bibitem{jantzen79} 
R. T. Jantzen, 
    Commun. Math. Phys. {\bf 64}, 211 (1979)  \vskip8pt  

\bibitem{wald84} 
R. M. Wald, 
    {\it General Relativity} 
    (University of Chicago Press, Chicago, 1984)  \vskip8pt   

\bibitem{batlle/al86} 
C. Batlle, J. Gomis, J. M. Pons, and N. Roman-Roy, 
    J. Math. Phys. (N.Y.) {\bf 27}, 2953 (1986)  \vskip8pt  

\bibitem{pons/shepley95} 
J. M. Pons and L. C. Shepley, 
    Class. Quantum Grav. {\bf 12}, 1771 (1995) gr-qc/9508052 \vskip8pt  

\bibitem{pons96} 
J. M. Pons, 
    Int. Jour. Mod. Phys. A {\bf 11}, 975 (1996)  \vskip8pt  

\bibitem{ryan/shepley75} 
M. P. Ryan and L. C. Shepley, 
    {\it Homogeneous Relativistic Cosmologies} 
    (Princeton University Press, Princeton, 1975)  \vskip8pt  

\bibitem{koike/tanimoto/hosoya94}
T. Koike, M. Tanimoto, and A. Hosoya,
    J.\ Math.\ Phys.\ {\bf 35},4855 (1994) gr-qc/9405052 \vskip8pt  

\bibitem{tanimoto/koike/hosoya97a}
T. Koike, M. Tanimoto, and A. Hosoya,
    J.\ Math.\ Phys.\ {\bf 38},350 (1997) gr-qc/9604056 \vskip8pt  

\bibitem{tanimoto/koike/hosoya97b}
T. Koike, M. Tanimoto, and A. Hosoya,
    gr-qc/9705052 \vskip8pt  

\bibitem{pons/salisbury/shepley96} 
J. M. Pons, D. C. Salisbury, and L. C. Shepley, 
    Phys. Rev. {\bf D55}, 658 (1997) gr-qc/9612037 \vskip8pt  

\bibitem{coussaert/henneaux93} 
O. Coussaert and M. Henneaux, 
    Class. Quantum Grav. {\bf 10}, 1607 (1993) gr-qc/9301001

\end{references}
\end{document}